\documentclass[3p,review,times]{elsarticle}
\biboptions{sort&compress}

\usepackage[T1]{fontenc}
\usepackage[utf8]{inputenc}
\usepackage{siunitx}
\usepackage[intlimits]{amsmath}
\usepackage{amssymb}
\usepackage{color}
\usepackage{pdfpages}

\usepackage{hyperref}

\newcommand{\tens}[1]{\boldsymbol{#1}} 
\newcommand{\tx}[1]{{\text{#1}}}

\renewcommand{\vec}[1]{\boldsymbol{#1}} 

\newcommand{\mean}[1]{\ifinner%
  \langle #1 \rangle%
  \else%
  \left\langle #1 \right\rangle%
  \fi}

\begin{document}

\begin{frontmatter}

\title{Light-induced deformation of side-chain azo-polymer:\\
Insights from atomistic modeling}

\author[ipf,tud]{Dmitry A. Ryndyk}
\author[ipf]{Olga Guskova}
\author[ipf,tud]{Marina Saphiannikova \corref{cor1}}
\ead{grenzer@ipfdd.de}

\cortext[cor1]{Corresponding author}
\address[ipf]{Division Theory of Polymers, Leibniz Institute of Polymer Research Dresden,
            01069 Dresden, Germany}
\address[tud]{Technische Universität Dresden, 01062 Dresden, Germany}

\begin{abstract}

In this study, we apply, for the first time, the fully atomistic force field approach to modeling light-induced deformations of azo-polymers, thereby establishing a relationship between macroscopic parameters and the microscopic molecular architecture of the used azo-polymers.
We apply an orientation potential to mimic the illumination of the sample, in which the action of light is represented through controlled redistribution of azo-chromophores relative to the polarization direction. This strategy allows us to capture both the microscopic details of chromophore behaviour and the collective, anisotropic response of the polymer matrix. By combining these complementary perspectives, the simulations not only resolve the local mechanism of light-induced motion but also provide a pathway to bridge molecular-scale dynamics with mesoscopic deformation phenomena in azo-polymer films.

\end{abstract}

\end{frontmatter}

\section{Introduction}
\label{sec:intro}

Superficial restructuring of photosensitive polymers has matured in recent years 
from the inscription of regular sinusoidal structures by the interference pattern of laser beam~\cite{Yadavalli2013,Yadavalli2013b,Jelken2019,Pagliusi2019,Priimagi2020,Lee2021} to the fabrication of intricate patterns by means of structured light~\cite{Oscurato2022, McGee2023}. Nowadays, the structure of laser beam can be easily reshaped in a lab to produce any desired pattern using such tools as geometric phase elements (q-plates) or spatial light modulators ~\cite{Marrucci2006,Forbes2021}. These tools enable 2D control of the intensity, polarization and phase of the propagating beam in the transverse plane, while light-matter interactions can introduce additional spatial dependence along the propagation direction. For example, photosensitive azobenzene-containing polymers (azo-polymers) not only absorb the light but can also rotate the azimuth of polarization ellipse clockwise or anti-clockwise depending on the helicity of the light and the penetration depth~\cite{Nikolova2000,Tverdokhleb2025}.

A unique property of side-chain azo-polymers is their ability to convert structured light irradiation into a well-defined stress field ~\cite{Azoreview2024, Oscurato2025} as a result of the light-induced reorientation of azobenzene chromophores (azo-chromo-phores). The stress magnitude in each material point is proportional to the local light intensity, while the major and minor axes of polarization ellipse define the principal axes of stress tensor ~\cite{Tverdokhleb2025}. Linearly polarized light causes a uniaxial stretching/contraction of azo-polymers ~\cite{Bublitz2000,Kang2014,Yadavalli2015,Loebner2018,Yadav2019}, while elliptically polarized light evokes biaxial deformation ~\cite{Toshchevikov2023,Tverdokhleb2025}. Implementing the light field into the finite element software ANSYS, an excellent agreement with experimentally observed superficial restructuring of azo-polymer films ~\cite{Tverdokhleb2023, Tverdokhleb2025} and reshaping of azo-polymer microobjects ~\cite{Yadav2019, Yadav2022, Oscurato2025} has been achieved. 

\begin{figure}[!t]
  \centering
	(a)\includegraphics[width=0.3\columnwidth]{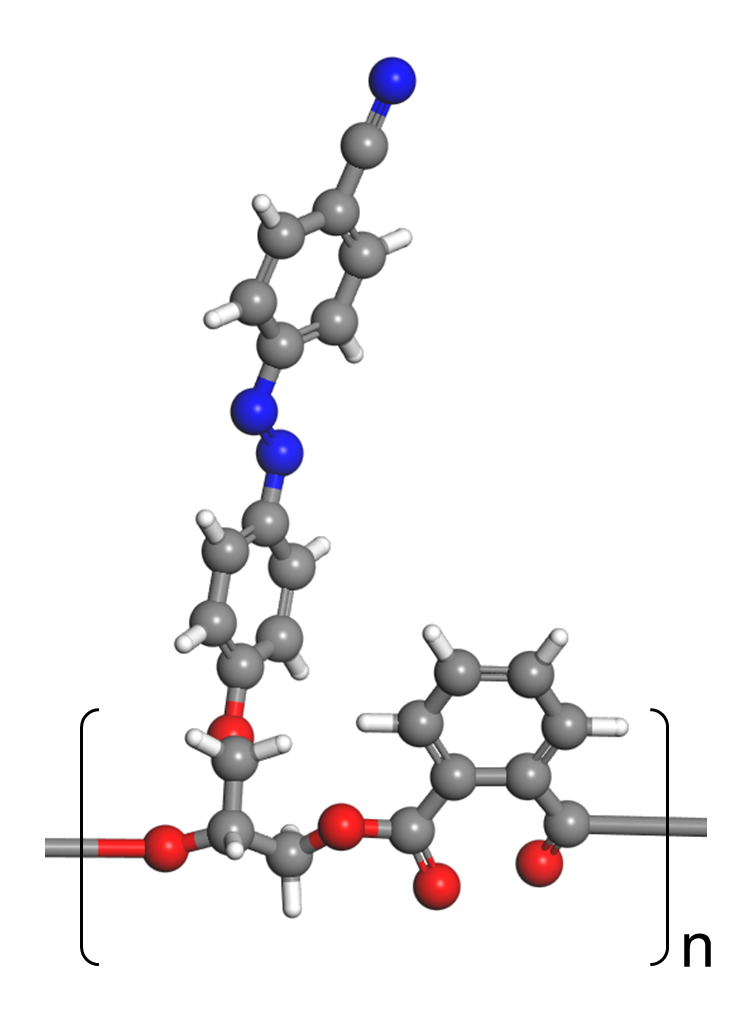} \hspace{2cm}
    (b)\includegraphics[width=0.2\columnwidth]{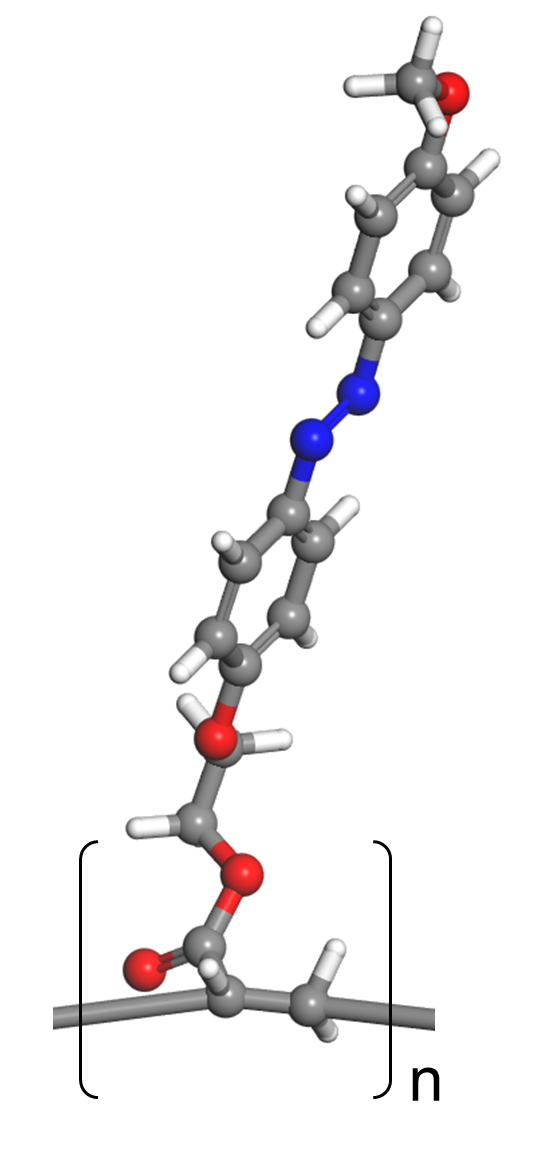}
	\caption{(a) The monomeric unit of azo-polyester with a stiff backbone and oxymethyl group as a spacer to azobenzene. (b) The monomeric unit of azo-polyacrylate with a flexible backbone and longer spacer to azobenzene consisting of ester- and ethyloxy group. }
	\label{azo-monomer}
\end{figure}

A long-term stability of the reshaped surfaces and objects is ensured by the use of
amorphous azo-polymers whose glass transition temperature is sufficiently high and is not affected by the light irradiation~\cite{Wu2019}. In such polymers the photosensitive azo-units are attached to the main chain via a short spacer.
Using computer simulations it has been verified that the orientation of azo-chromophores perpendicular to the polarization direction is transferred to the reorientation of polymer backbones along the same direction~\cite{Ilnytskyi2011,Ilnytskyi2019}. The models of amorphous azo-oligomers in the simulations were coarse-grained (azo-unit being represented either as Gay-Berne model (GB) or spherocylinder particle) and rather stiff. Not only torsional barriers were increased both in backbones and side-chains, but also a single spherical bead represented the spacer. The use of stiff models was motivated by the experiments of Bublitz et al.~\cite{Bublitz2000}, who studied light-induced elongation of azo-polyester droplets on the water surface. The monomeric unit of this stiff azo-polyester is shown in Fig.~\ref{azo-monomer}a. In the simulations~\cite{Ilnytskyi2011}, the reorientation of the azo-oligomers was accompanied by a significant extension of the volume element along the polarization direction, reproducing the experimentally observed effect~\cite{Bublitz2000}. Due to a strong (fast rigid) coupling of side- and main chains, the latter were oriented under light irradiation on the same time scale as the azo-chromophores.

Atomistic molecular dynamics simulations that explicitly include photoisomerization of azo-units have also been employed to study reshaping of thin glassy azo-polymer films.~\cite{bockmann2016towards,merkel2023understanding} These simulations provide a microscopic description of how repeated trans-cis-trans cycles of the chromophores can trigger molecular migration and surface reshaping. However, due to the direct deposition of photon energy into the local degrees of freedom, such simulations are accompanied by pronounced transient heating in the irradiated regions, with temperatures reported up to 1800~K~\cite{bockmann2016towards}. While these temperatures exceed the stability range of organic materials and therefore cannot be directly compared with experiment, they enable the mass transport process to be observed on the nanosecond timescales accessible to molecular dynamics of a side-chain azo-polymer. In parallel, theoretical approaches based on the reorientation of azo-chromophores~\cite{Toshchevikov2014,Toshchevikov2017,Yadav2022}, describe the light-induced mass transport without invoking elevated temperatures. These models attribute the effect to anisotropic stresses arising from orientational ordering of polymer backbones and thus reproduce directional photodeformations and surface relief grating formation under near-ambient conditions.

Various azo-polymers have been utilized in the labs for superficial restructuring of thin films or for reshaping of micropillar arrays. Oft polymers with flexible backbones and a spacer containing longer alkyl or alkyloxy groups have been used. 
For example, the spiral patterns in the experiment with helical beams were produced on the amorphous films of acrylic polymer with the methoxyazobenzene side-chain~\cite{Ambrosio2012}, see Fig.~\ref{azo-monomer}b.
The same polymer has been recently used to demonstrate the insciption capabilities of computer-generated holography~\cite{Oscurato2019, Oscurato2023}. For such a flexible polymer the coupling between the orientation of azo-containing side-chains and the backbones should be much weaker in comparison to the stiff azo-polyester discussed above. Nevertheless, the superficial restructuring is shown to be rather effective. This can only mean that this polymer also effectively converts the structured light irradiation into the stress field which causes directional photodeformations.

To clarify the nature of the azo--backbone coupling, we perform in this study fully atomistic molecular dynamics MD simulations of dense azo-polyacrylate samples in a wide range around their glass transition temperature. To speed up the simulations, the alignment of azos perpendicular to the polarization direction is enforced by the light-induced orientation potential that arises naturally from multiple cycles of photoisomerization~\cite{Toshchevikov2017,Yadav2019}.  The main questions on which we would like to receive the answers are: 1) Is it possible to observe the orientation or stretching of acrylate backbones along the polarization direction in fully atomistic simulations?   2) If so, what parameters would define the separation of time scales between the orientation of azos and the stretching of backbones? 3) Does the length of the main chains matter?
We invite a fellow reader to follow us on this quest, which will hopefully provide valuable insights into the nature of polarized light-induced processes at the microscopic scale.

\section{Fully atomistic model}
\label{sec:simu}

\subsection{Preparation of amorphous azo-polymer}

The all-atom model of azo-monomer is built in Materials Studio 9.0~\cite{MatStud}. The monomeric unit has acrylate backbone and a side-chain composed of 4-methoxyazobenzene grafted to a backbone at azobenzene's $4'$-position via ethyloxy group, as depicted in Fig.~\ref{azo-monomer}b. The geometry of the monomeric unit is optimized for the trans-isomer of azobenzene group, using Forcite module with the following settings: polymer-consistent force field (\textit{PCFF})~\cite{sun1994ab}, \textit{Smart} optimization algorithm (energy convergence 2$\times10^{-5}$ kcal/mol, force convergence tolerance 0.001 kcal/mol/$\AA$, displacement convergence 10$^{-5}$ $\AA$). 

This optimized geometry is further used for "polymerization" procedure in AMBER software~\cite{case2023ambertools}. Two compositions are modeled:  Oligomers with chain lengths of $N_{mol}=10$ and 20 monomeric units and  with number of chains in the simulation box $n=96$ and 48, respectively. Both oligomers are constructed in a head-to-tail configuration and exhibit atactic stereochemistry. The general AMBER force field (\textit{GAFF})~\cite{wang2004development} is applied for the parametrization of the bonded and non-bonded interactions in the simulation compositions. This force field has been validated for the simulations of various azo-containing materials ~\cite{moghaddam2020theoretical,ehrman2021improving,heinz2008photoisomerization}. The most important parameters for our simulation to take into account are (i) the parameters for the torsion angle of the azo group preventing the trans-cis transition and keeping this mesogene group planar and (ii) the dihedrals for the proper rotation of the side chains around the bond of the attachment to the main chain. The first one is taken to be artificially large. The reason is that we assume the frequency of light keeps the azobenzene in trans-state almost all the time and produces rotational potential due to virtual transitions. The latter defines the strength of the "side-chain--backbone" coupling, i.e. the ability to transmit the torque acting on the azo-group to the main polymer chain, as given within the concept of the orientation approach ~\cite{Toshchevikov2009,Toshchevikov2014,Toshchevikov2017,Toshchevikov2023}. 

Initially, in the cubic periodic simulation boxes with \textit{L}=150 $\AA$ for $N_{mol}=10$ and \textit{L}=200 $\AA$ for $N_{mol}=20$, the stretched polymer chains are arranged in an array with their backbones oriented parallel to each other. Using first NVT molecular dynamics simulations implemented in LAMMPS~\cite{thompson2022lammps}, the systems were equilibrated to an amorphous state~\cite{anstine2020effects} at $T = 1000$ K (Nose-Hoover thermostat). Then we applied NPT molecular dynamics simulations at normal pressure $P$ = 1 atm $\approx 10^{-4}$ GPa (Nose-Hoover barostat) over a simulation time of 10 ns with a time step of 1 fs. Non-bonded van der Waals interactions are truncated at a cutoff distance of 12 \AA, while long-range electrostatic interactions are computed using the particle-particle particle-mesh (PPPM) summation technique. Partial atomic charges are assigned based on the restrained electrostatic potential (RESP) fitting method~\cite{bayly1993well}. The resulting density at $T=1000$ K is about 0.76 g/cm$^3$ for both compositions.

\subsection{Glass transition temperature}

\begin{figure}[!h]	\centering\includegraphics[width=0.8\columnwidth]{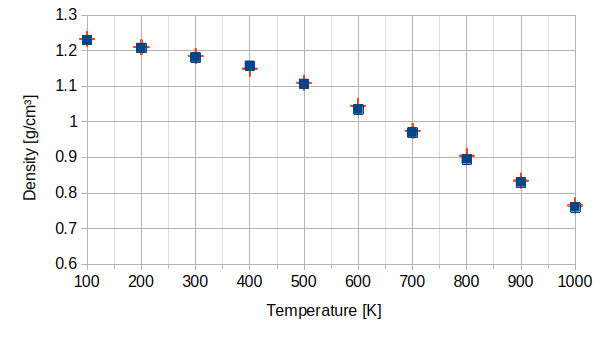}
	\caption{\textit Determination of \textit{T}$_g$ $\approx$ 450 K  for two simulated chain lengths $N_{mol}=10$ (blue squares) and $20$ (red crosses).}
	\label{Tg}
\end{figure}

Next, the glass transition temperature should be determined. From experiments of Ambrosio et al.~\cite{Ambrosio2012} the glass transition temperature for the same polymer with $N_{mol}=86$ monomeric units is $T_g = 67$°C. Above this temperature the sample is in nematic state until it isotropizes at $T_{iso}=113$°C. 
The most common method for determining glass transition temperature in MD simulations involves monitoring the density of the polymer as a function of temperature \cite{Ilnytskyi2011}. Here, NPT simulations are performed over a range of temperatures, starting from a high-temperature equilibrium state (1000 K) down to cryogenic temperatures (100 K), with stepwise cooling (100 K step).  At each temperature, the system is equilibrated for at least 10 ns. The transition from a fluid-like to a glassy state is identified by a change in the slope of the density-temperature curve, presented in Fig.\,\ref{Tg}. The glass transition temperature is observed to be  approximately 450 K. The predicted \textit{T}$_g$ exhibits only marginal dependence on the molecular mass, consistent with recent findings from systematic all-atom simulations~\cite{klajmon2023glass}. Interestingly, this all-atom \textit{T}$_g$ prediction is in excellent agreement with earlier coarse-grained simulations, which reported \textit{T}$_g$ $\approx$ 445 K for the same polymer architecture \cite{Ilnytskyi2011,saphiannikova2013nanoscopic}. This remarkable consistency across different levels of resolution confirms the robustness of the simulated \textit{T}$_g$ and highlights that both modeling approaches capture the essential role of azo side groups and backbone rigidity in the glass transition. The agreement between coarse-grained and all-atom models thus provides strong support for the transferability of simulation results and suggests that simplified models can reliably reproduce key thermophysical properties such as \textit{T}$_g$.

It is worth noting that the simulated \textit{T}$_g$ of about 450 K significantly exceeds the experimental value reported by Ambrosio et al.~\cite{Ambrosio2012}. Such a discrepancy is a well-known feature of fully atomistic simulations and may be attributed to several factors. First, the cooling rates accessible in MD are orders of magnitude faster than those in experiment, leading to a systematic upward shift of \textit{T}$_g$. Second, limitations in the force-field parametrization, particularly in describing subtle balances of non-bonded interactions and torsional potentials, can bias the absolute position of the transition. 
Taken together, these factors highlight that the simulated Tg should be interpreted in a relative rather than absolute sense, while still offering a reliable framework for understanding the underlying molecular mechanisms.


\subsection{Orientation potential}

The kinetics of the photoisomerization process
and light-induced ordering in azo-containing materials have been studied in detail in the Refs.~\cite{Toshchevikov2017} and ~\cite{Petrova2017}. Here, we briefly describe the key features that allow the introduction of an effective orientation potential, the use of which significantly reduces the computational effort. It is well known that the angle-selective absorption of photons by the trans-isomers causes an orientation of their long axes in a direction perpendicular to the light polarization $\vec E$. The stationary state is reached after tens of photo-isomerization cycles~\cite{Petrova2017}. Although the absorption of photons by the cis-isomers is not angular-selective, their population becomes enriched around $\vec E$ due to the depletion of a number of trans-isomers in the same direction. The orientational states of trans- and cis-populations have been described by the two order parameters $S_T$ and $S_C$ in respect to $\vec E$ and there relative fractions $\Phi_T$ and $\Phi_C= 1-\Phi_T$. 
The time evolution of these four parameters and their stationary values strongly depend on the ratio $\tilde P_C=P_C/P_T$ between the probabilities of cis-trans and trans-cis isomerization. The average order parameter $\bar S=\Phi_T S_T + \Phi_C S_C$ for all chromophores is shown to be nearly zero for short wavelengths ($\lambda \leq 400$ nm), when $\tilde P_C$ falls to 0.1 and $\Phi_C \geq 0.9$. Contrary, at longer wavelengths ($\lambda \geq 480$ nm), when $\tilde P_C$ increases to 10 and $\Phi_C \leq 0.1$, the average order parameter has large negative values. 
More importantly, its time evolution can be
very well described with the help of effective orientation potential
     \begin{equation}
         U=V_0 (\vec u\cdot \hat{\vec E})^2=V_0 \cos^2\theta,
    \label{eff_potential}     
    \end{equation}
where $\vec u$ is the unit vector of azobenzene orientation, $\hat{\vec E}$ is the unit vector of light polarization and $\theta$ is the angle between these two vectors. The strength of potential $V_0 \sim kT P_T/D$ is defined by the probability of trans-cis isomerization and the orientation diffusion coefficient of the azobenzene $D$. The ratio  $kT/D$ is proportional to the material viscosity $\eta$ which is known to increase exponentially with the decrease of absolute temperature $T$ and reach the value of $10^3$ $\text{GPa}\cdot\text{s}$ at the glass transition temperature $T_g$. This implies that the strength of potential should increase upon cooling the material at the same illumination intensity. 

The effective orientation potential  \eqref{eff_potential} exerts a torque in the azimuthal direction $\hat{\vec \phi}$
    \begin{equation}
         \vec M = -\vec u \times \nabla_{\vec u} U = V_0 \sin 2\theta \hat{\vec \phi},
    \label{torque}     
    \end{equation}
that acts collectively on all atoms of the rigid azo-unit, causing them to rotate in concert. As can be seen from Eq.~\eqref{torque}, torque is measured in the same units as potential energy and has a maximum magnitude of $V_0$ when $\theta = 45^{\circ}$. In this study, we applied the light-induced torques with the values of $V_0$ ranging from $5$ to $40$ kcal/mol. These values correspond to $3.5 - 27.8$ $10^{-20}$ J and are one order of magnitude higher than those used in coarse-grained modeling of amorphous azo-polymers \cite{Ilnytskyi2011}. There, the sample density was about half as high, which probably explains lower values of the applied torques. 

The magnitude of light-induced stress acting on azo-chromophores in the beginning of irradiation can be estimated as follows \cite{Toshchevikov2017,Yadav2022}:
     \begin{equation}
        \tau_0 = 2nV_0/5
    \label{stress}    
    \end{equation}
where $n$ is the number density of chromophores. It slightly increases with the decrease of temperature due to a higher density of the simulated box. Let's choose $T=550$ K, at which the box has the edges of about 80 $\AA$. There are 960 azo-chromophores in the box. This gives $n=1.875 \cdot 10^{27}$ m$^{-3}$, which is very close to the value of $1.5 \cdot 10^{27}$ m$^{-3}$ used in our previous theoretical studies \cite{Toshchevikov2009,  Toshchevikov2017}. For the light-induced torque $V_0 = 10$ kcal/mol $\approx 6.95 \cdot 10^{-20}$ J, the estimated stress magnitude is $130$ MPa. Such a high stress is required to observe the reorientation of chromophores within a few tens of nanoseconds, which are accessible in fully atomistic MD simulations.

\section{Light-induced effects at different temperatures and light intensities}

\subsection{The order and shape parameters}

After equilibration at different temperatures and normal pressure, we switch on the orientation potential and analyze the real-time behavior of light-induced orientation and deformation in azo-polyacrylate samples. The orientational order parameter for azo-chromophores
     \begin{equation}
        S_\tx{azo}=\left<\frac{3}{2}(\vec u\cdot \vec n)^2-\frac{1}{2}\right>
    \label{Azo-order}    
    \end{equation}
is defined in respect to the spontaneous nematic director $\vec n$. It aligns along the polarization direction of light under the action of orientation potential.  
The angular brackets designate an averaging over all chromophores, orientation of which is characterized by the unit vectors $\vec u$. The orientation order of oligomer backbones is described by the order parameters $S_\tx i$ calculated in respect to the coordinate axes $\vec i = \vec x, \vec y$ and $\vec z$:
    \begin{equation}
    \label{S_i}
    S_\tx{i}=\left<\frac{3}{2}(\vec b\cdot \vec i)^2-\frac{1}{2}\right>
    \end{equation}
The averaging is carried out over all oligomer backbones, whose orientation is characterized by the unit vectors $\vec b$. 

The directional photodeformations are monitored following dimensions of the volume element $L_x,L_y,L_z$, i.e., the lengths of the periodic simulation box, which evolve self-consistently under constant-pressure conditions (NPT ensemble). The shape of each backbone is described by
the eigenvalues and eigenvectors of its gyration tensor. 
Following our previous publication \cite{saphiannikova2013nanoscopic}, the orientation distribution of azo-chromophores around the oligomer backbones is characterized by two angles. The polar angle $\alpha = \arccos |\vec u \cdot \vec g_{max}|$ is measured between the orientation vector $\vec u$ of
the chromophore and the main eigenvector $\vec g_{max}$ of the gyration tensor. Only the smallest angle is chosen, since the orientation states with $\alpha$ and $180^{\circ} -\alpha$ should be equiprobable for side-chain azo-polymers \cite{Toshchevikov2009}. The azimuthal angle $\beta$ is measured within the plane formed
by the medium and minor eigenvectors.

\begin{figure}[h!]
    \begin{center} 
    (a) \includegraphics[width=0.6\columnwidth]{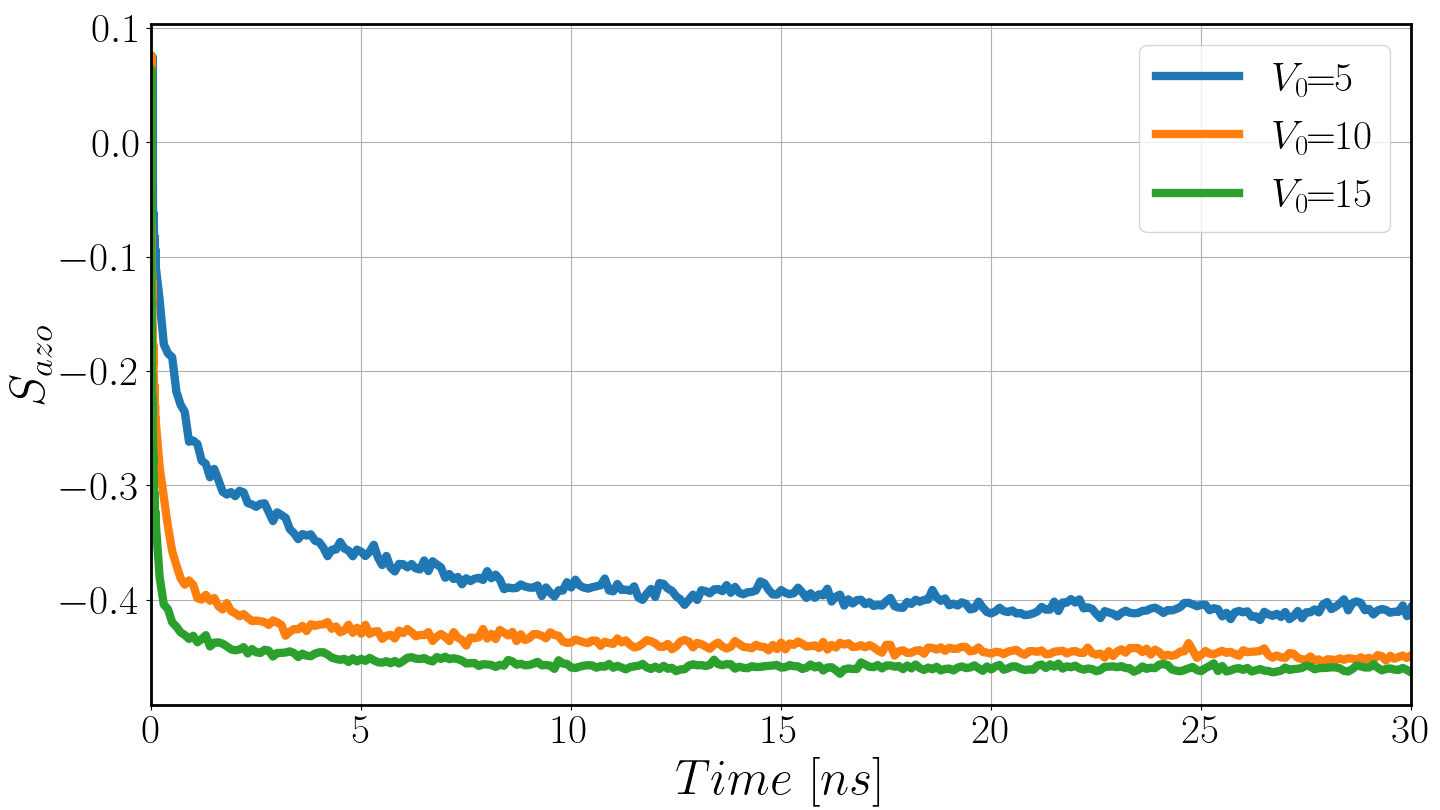}\\
    (b)
    \includegraphics[width=0.6\columnwidth]{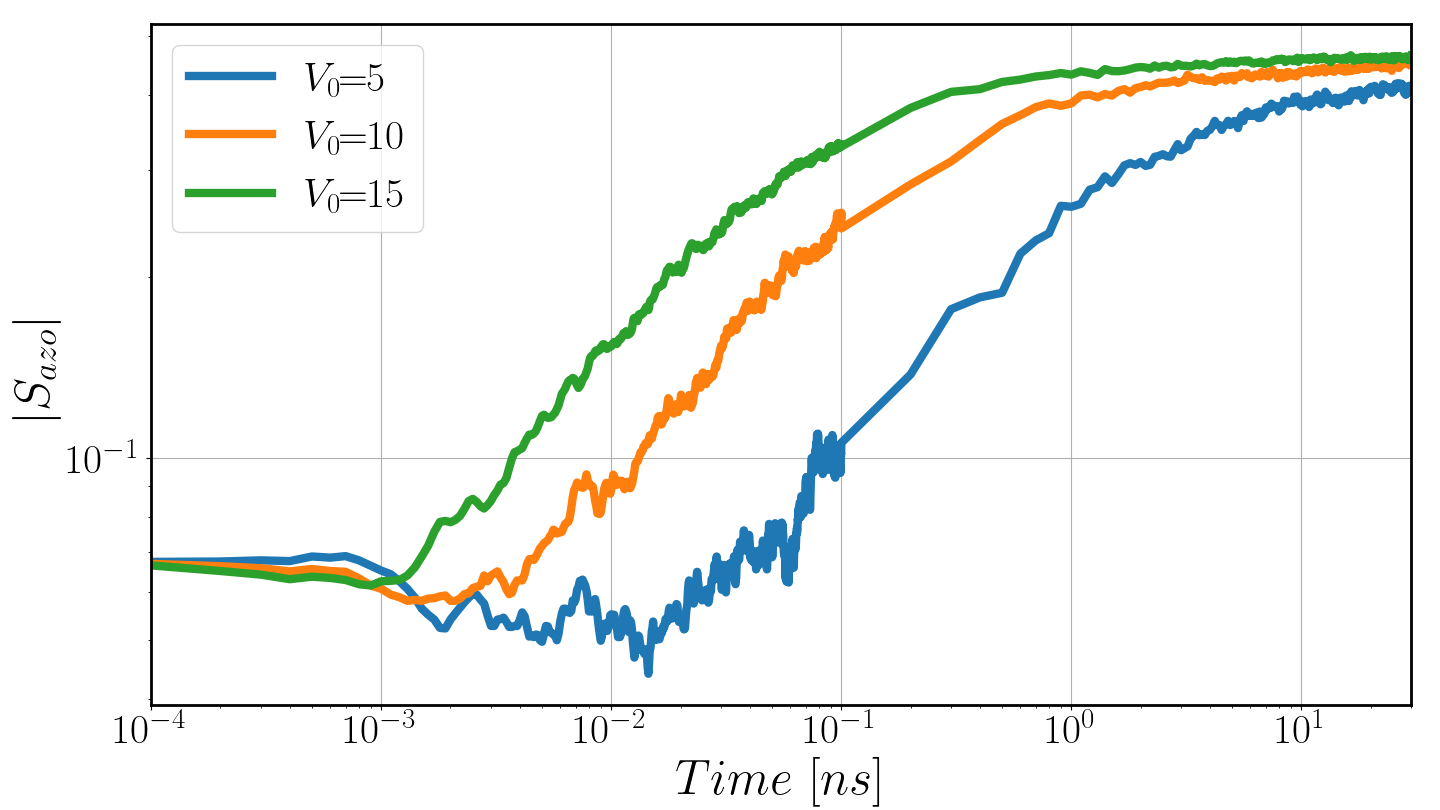}
    \end{center}
    \caption{(a) Orientational order parameter $S_{azo}$ for $N_{mol}=20$ at $T=550$ K and different light-induced torques $V_0$. (b) The double-logarithmic plot shows that time evolution of the order parameter magnitude $|S_{azo}|$ exhibits three regimes : 1) initial delay in reorientation, 2) exponential growth, 3) slow approach to the steady state. 
    }
    \label{S_azo20_T550}
\end{figure}

\subsection{Reorientation of azobenzene chromophores}
\label{sec:azo-orientation}

We begin by analyzing the orientational order parameter of chromophores, $S_{azo}$. In the absence of light, $S_{azo}$ fluctuates around zero, which is a typical behavior for isotropic systems of nanoscopic size \cite{Ilnytskyi2011}. 
After the light is switched on, the azo-chromophores reorient perpendicular to the light polarization $\vec E$, that breaks the symmetry of initially isotropic system. The spontaneous nematic director in Eq.~\eqref{Azo-order} aligns along the polarization direction, $\vec n = \hat{\vec E}$. Note that such a nematic state is characterized by negative values of the order parameter  $-0.5 \le S_{azo} < 0$. Therefore, $S_{azo}$ decreases monotonically with time under light irradiation and finally reaches the steady state plateau, whose value depends on the light intensity and temperature. At $T=550$ K, the plateau value for light-induced torques $V_0\ge 5$ kcal/mol is below $-0.4$, indicating a strong reorientation into the plane perpendicular to $\vec E$ (Fig.\,\ref{S_azo20_T550}a).

\begin{figure}[t]
    \begin{center}
    (a) \includegraphics[width=0.6\columnwidth]{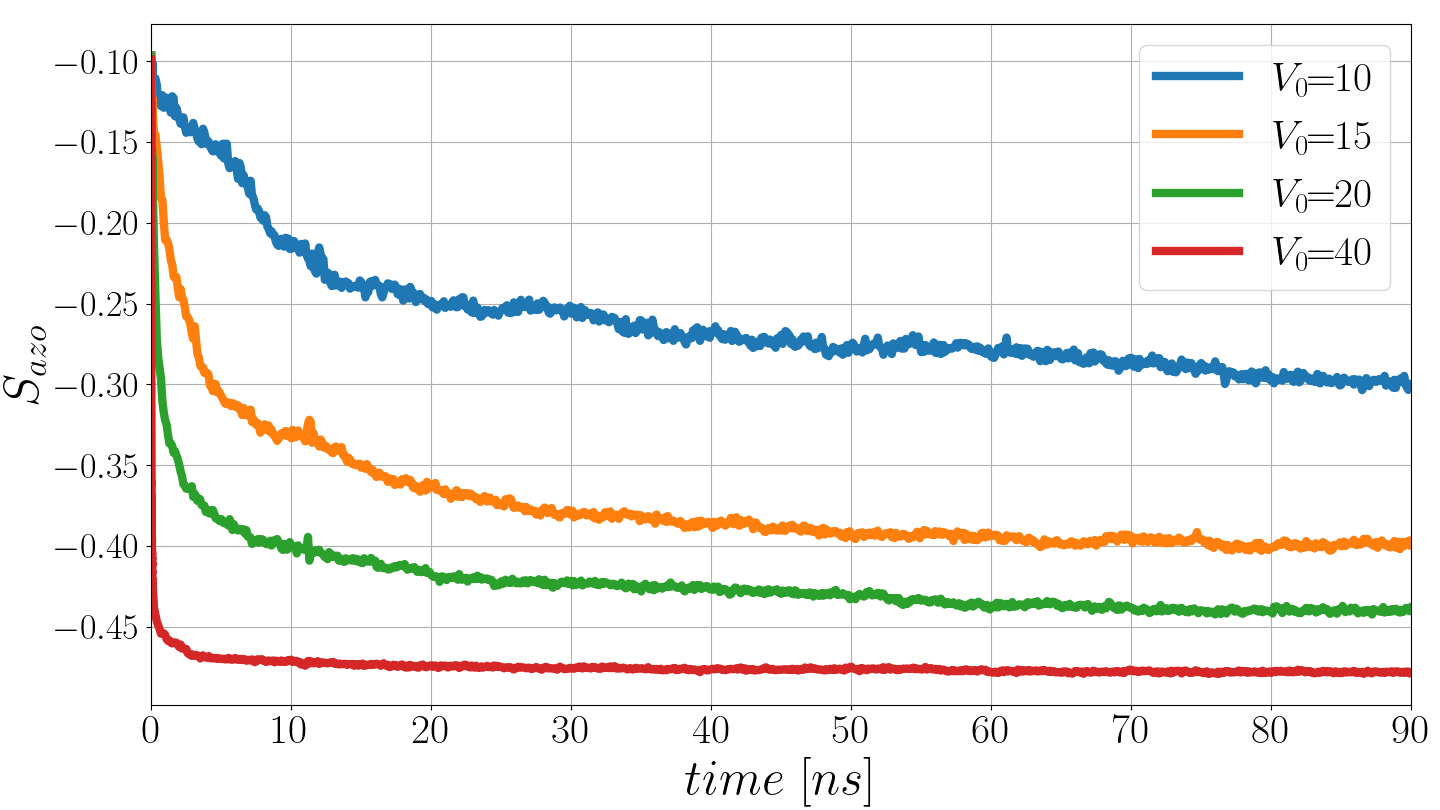}\\
    (b) \includegraphics[width=0.6\columnwidth]{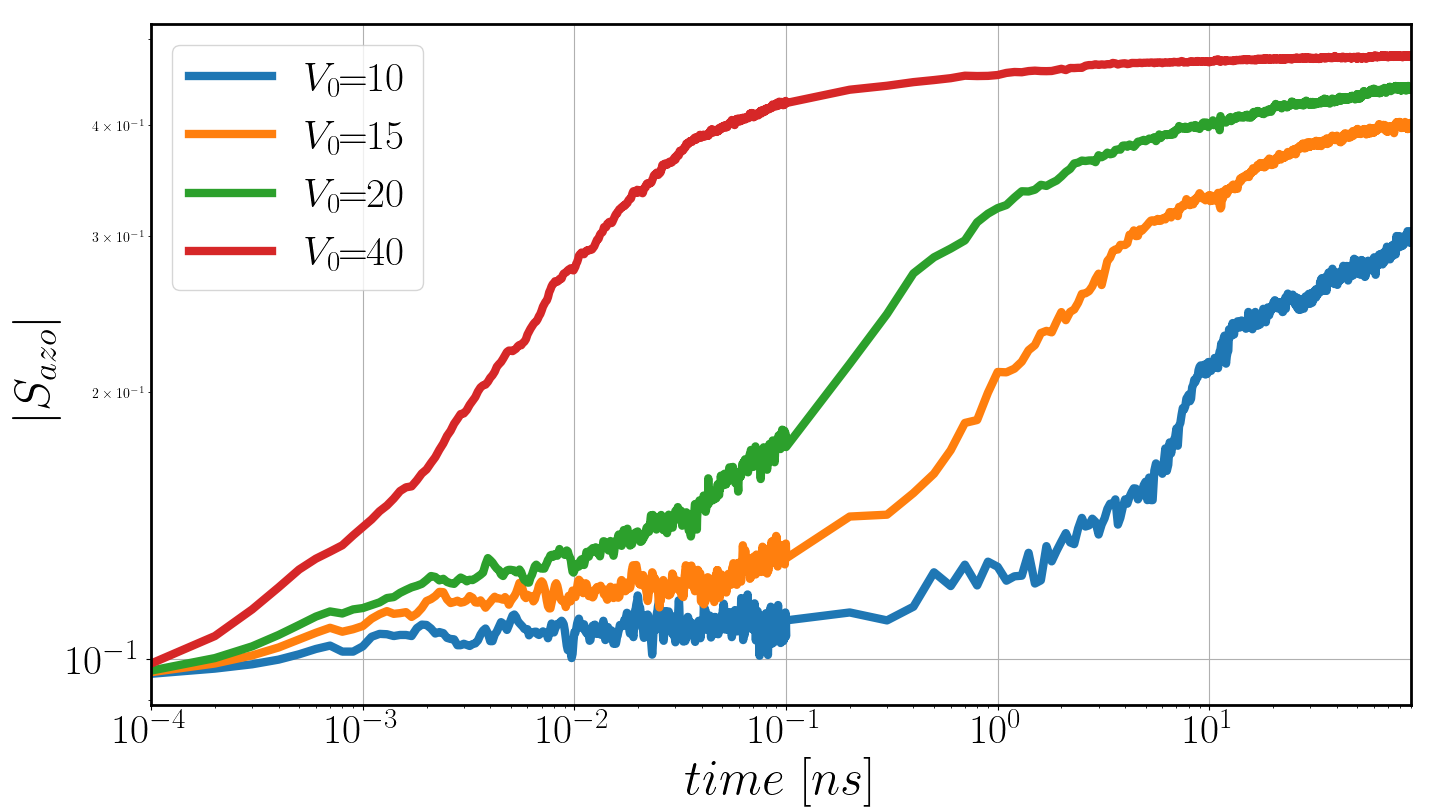}
    \end{center}
\caption{(a) Orientational order parameter $S_{azo}$ for $N_{mol}=20$ at $T=400$ and different light-induced torques $V_0$. (b) The double-logarithmic plot shows that time evolution of the order parameter magnitude $|S_{azo}|$ exhibits three regimes: 1) slow exponential growth, 2) fast exponential growth, 3) slow approach to the steady state. }
    \label{S_azo20_T400}
\end{figure}

Our simulations clearly show several timescales. For $T>T_g$, one can distinguish at least three different regimes of time evolution in the double logarithmic representation. As an example, we show in Fig.\,\ref{S_azo20_T550}b the simulations results at $T=550$ K. The first time $t_1$ characterizes the delayed reaction of single azo-chromophores on the orientation potential. It decreases with the increase of light-induced torque: $t_1$ is about $0.01$ ns for $V_0=5$ kcal/mol and $0.001$ ns for $V_0=15$ kcal/mol. At $t_1<t<t_2$, the azo-chromophores reorient collectively which is reflected by exponential growth of the order parameter magnitude $|S_{azo}|$. Finally, at $t > t_2$, the order parameter slowly approaches the steady state value. Since the time $t_2$ also decreases with the increase of $V_0$, the regime of exponential growth stretches approximately over the same two decades for all values of $V_0$. 

The similar behavior for time evolution of $S_{azo}$ is observed at $T<T_g$ but at larger values of the light-induced torque $V_0$. As an example, we show the simulations results at $T=400$ K (Fig.\,\ref{S_azo20_T400}). To achieve the same degree of reorientation, for instance $S_{azo}=-0.4$, it is necessary to apply $V_0 = 15$ kcal/mol, three times larger torque than at $T=550$ K (compare with Fig.\,\ref{S_azo20_T550}). Interestingly, at low temperatures, there is no initial delay in the orientation process: the order parameter magnitude begins to increase immediately after application of the light-induced torque. However, the reorientation takes much longer time, which is clearly evident at $V_0=10$ kcal/mol, where the steady state is not yet reached after 90 ns.     

\begin{figure}[t]
    \begin{center}
    (a) \includegraphics[width=0.6\columnwidth]{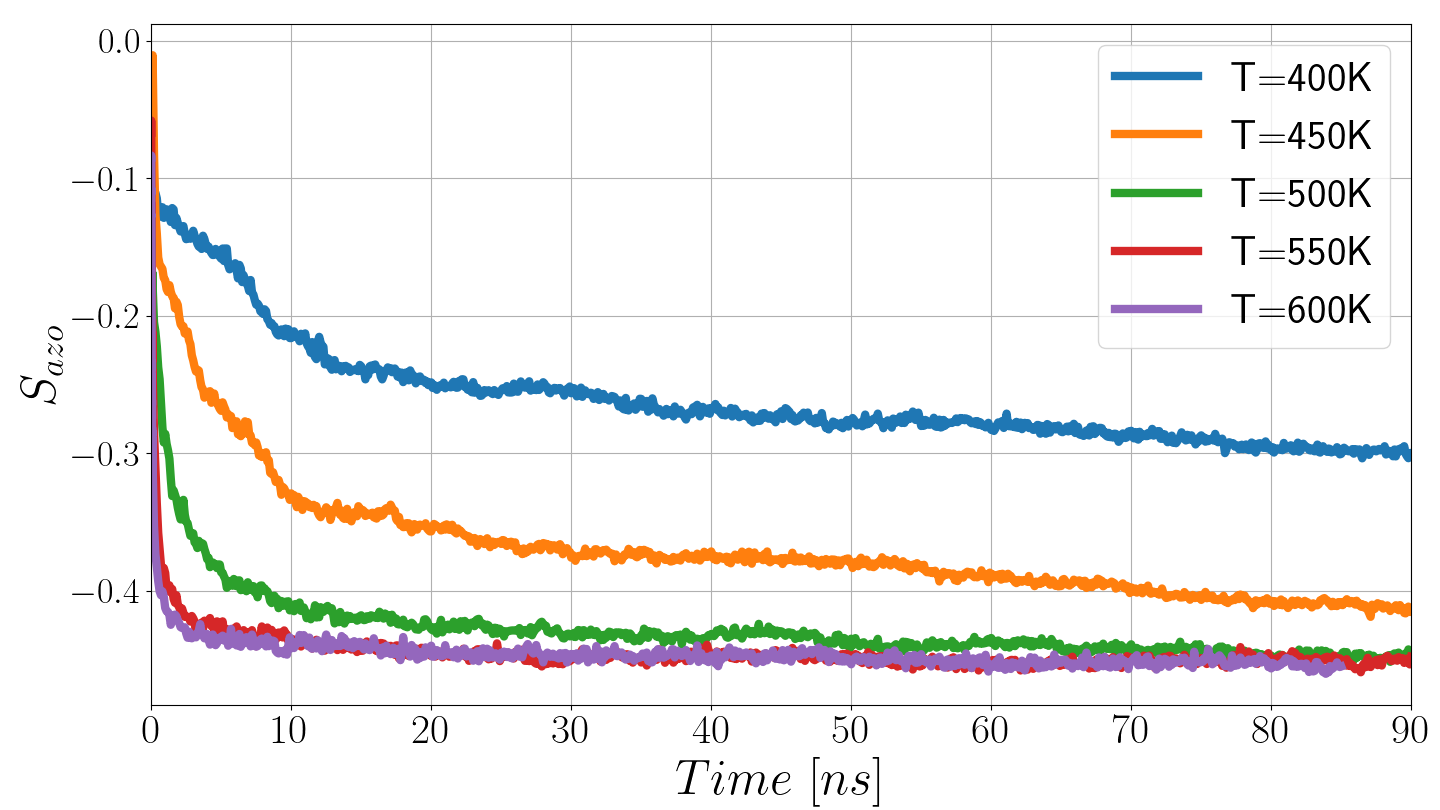}\\
    (b) \includegraphics[width=0.6\columnwidth]{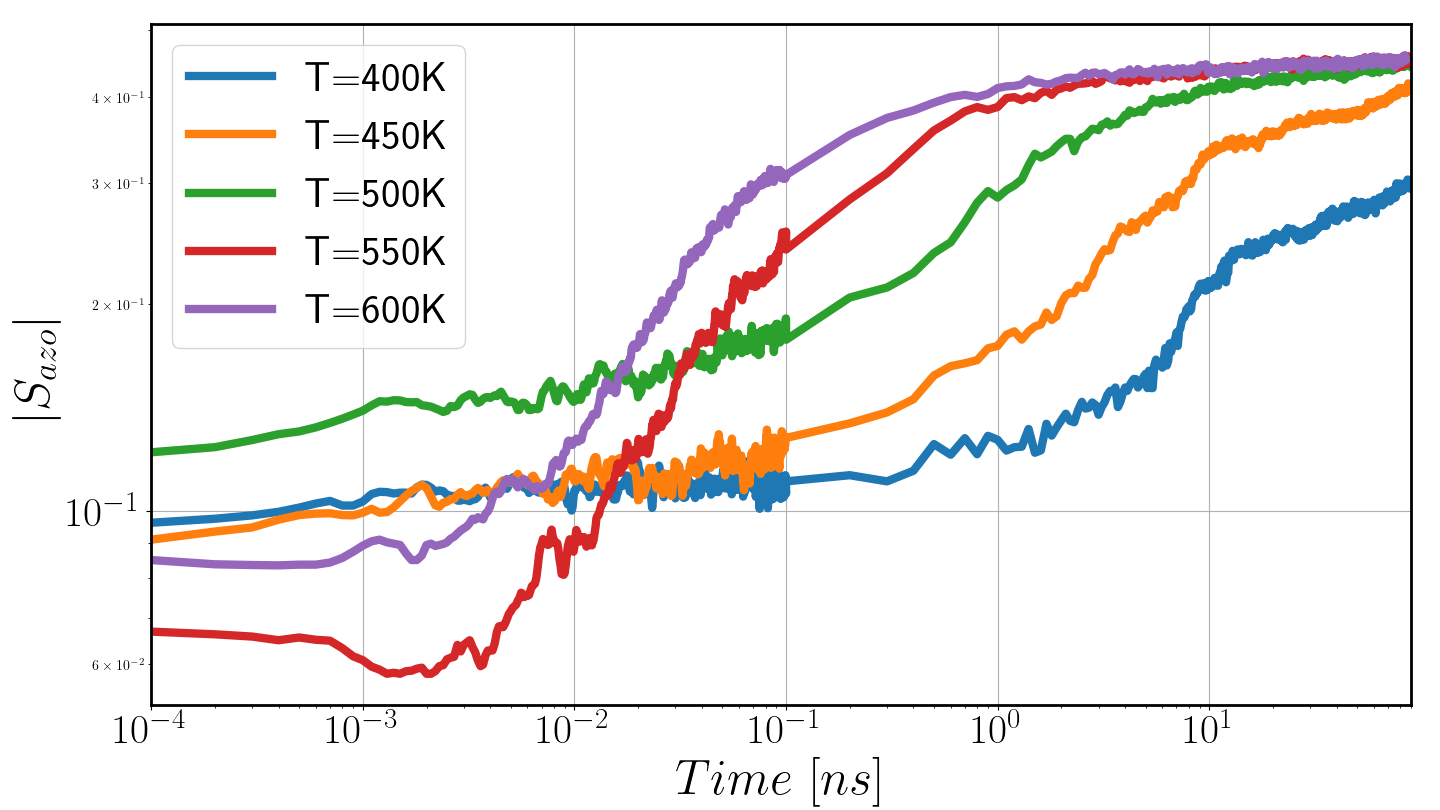}
    \end{center}
    \caption{(a) Orientational order parameter $S_{azo}$ for $N_{mol}=20$ at $V_0=10$ kcal/mol and different temperatures $T$. (b) The logarithmic plot shows the magnitude of order parameter $\ln|S_{azo}|$.}
    \label{S_azo20_F10}
\end{figure}

It is instructive to compare the reorientation of azo-chromophores at the same $V_0$ but different temperatures (Fig.\,\ref{S_azo20_F10}). Reorientation occurs very rapidly ($\sim 10$ ns) at all three temperatures above $T_g$ and starts to slows down at $T_g$. Nevertheless, the order parameter appears to approach the same steady-state value of $-0.45$.

\subsection{Directional photodeformations}
\label{sec:photodeformation}

Linearly polarized light causes uniaxial stretching of azo-polymer microposts \cite{Kang2014,Yadav2019} and colloids\cite{Loebner2018,Yadav2022} along the polarization direction, while preserving their volume.
The same effect can be investigated in fully atomistic MD simulations by tracking the temporal evolution of the dimensions of a nanoscopic sample. Irradiation of the samples with the light linearly polarized along the x-, y- and z-axes showed no differences at any temperature. This further confirms the isotropic orientation of the azos and polymer backbones in as-prepared samples. In the following we present the results of irradiation with light polarized along the x-direction. 

\begin{figure}[!t]
  \centering
\includegraphics[width=0.3\columnwidth]{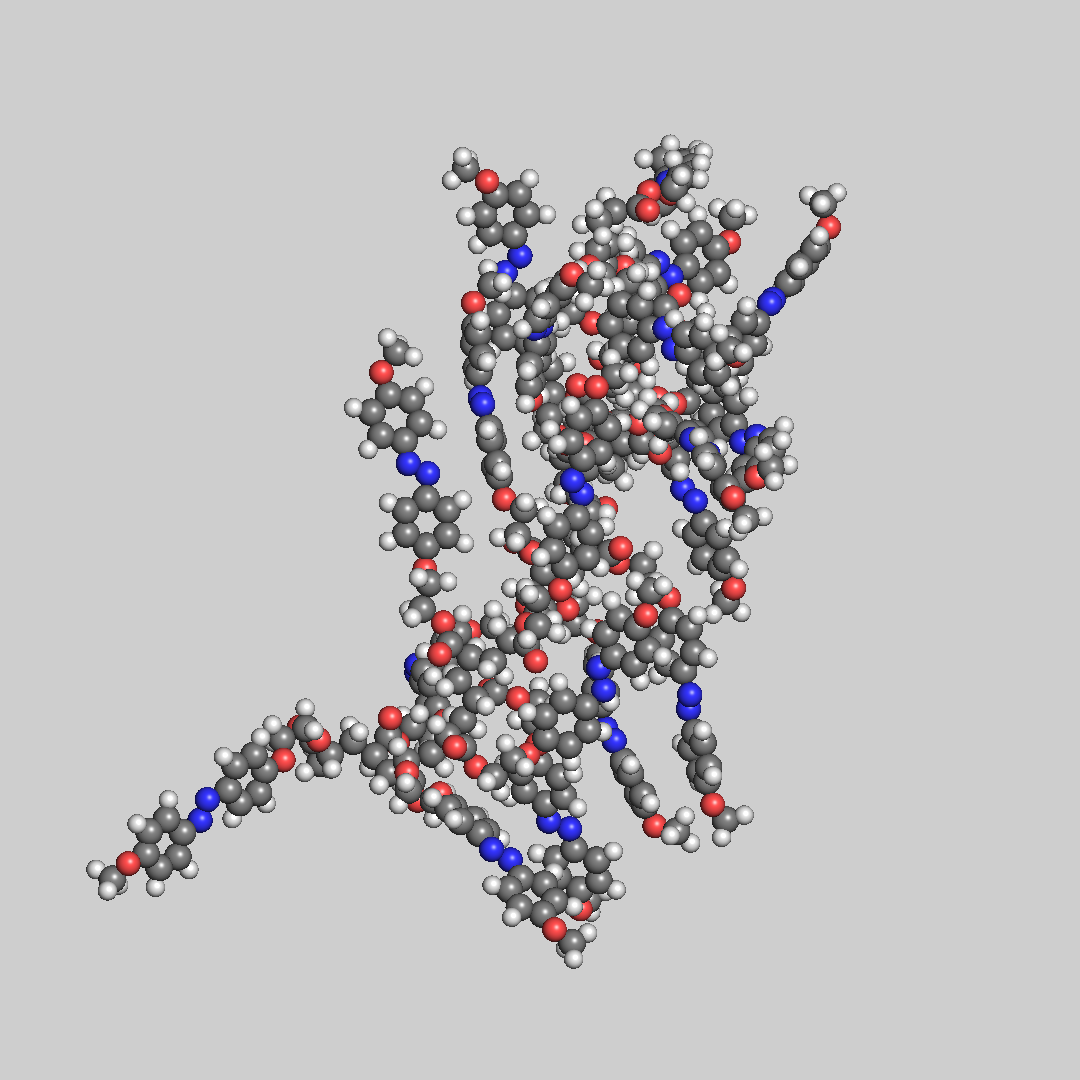}
\includegraphics[width=0.3\columnwidth]{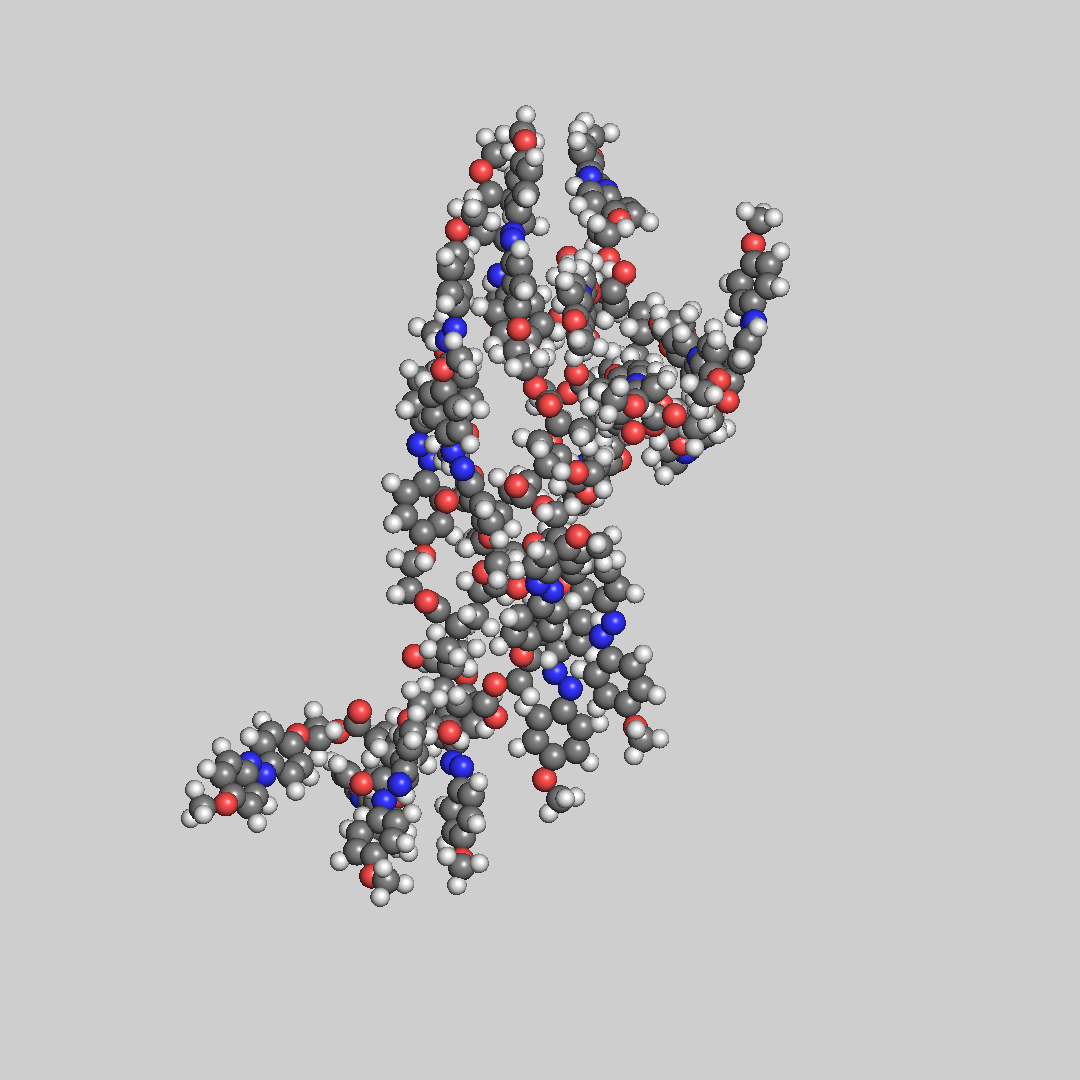}
\includegraphics[width=0.3\columnwidth]{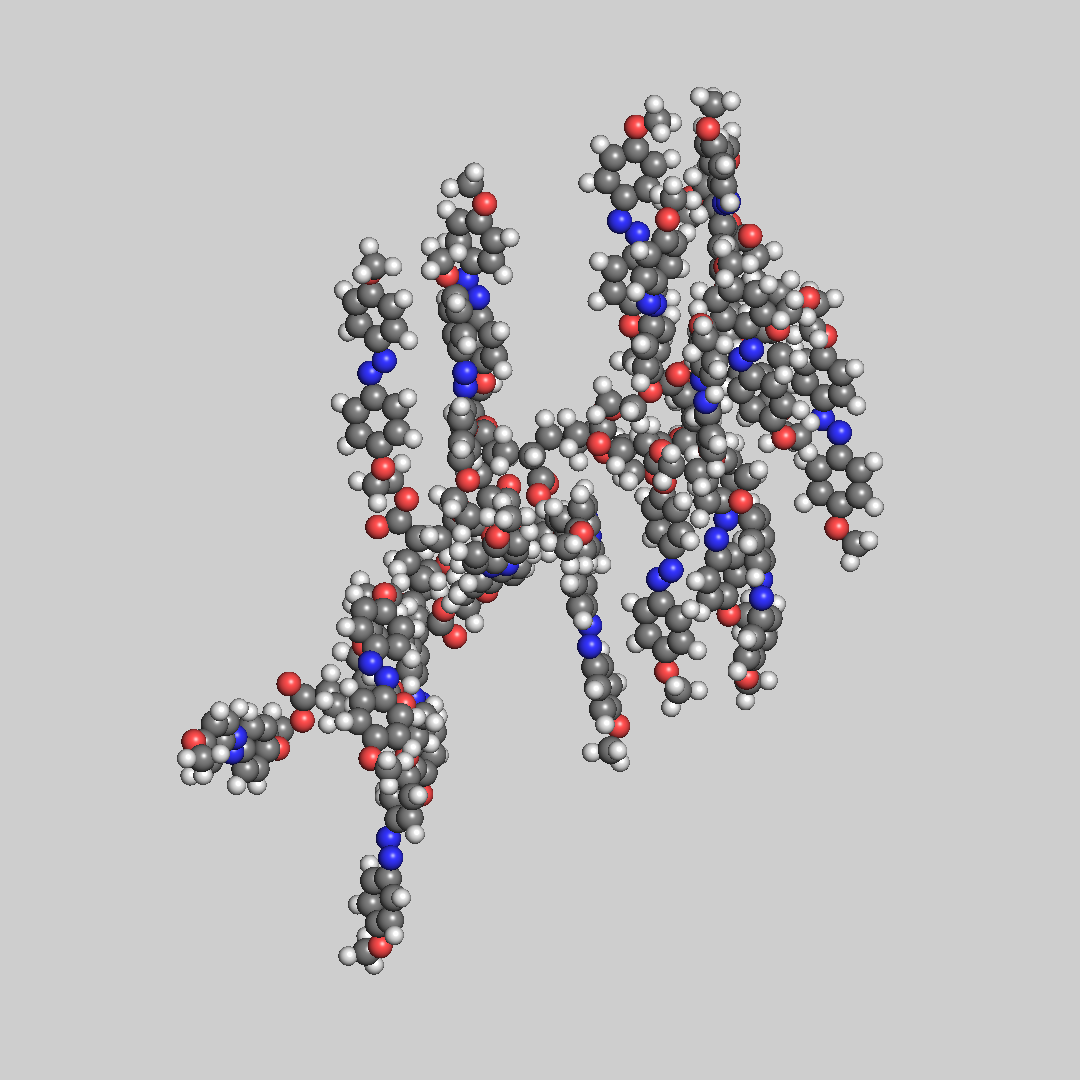}\\
\includegraphics[width=0.25\columnwidth]{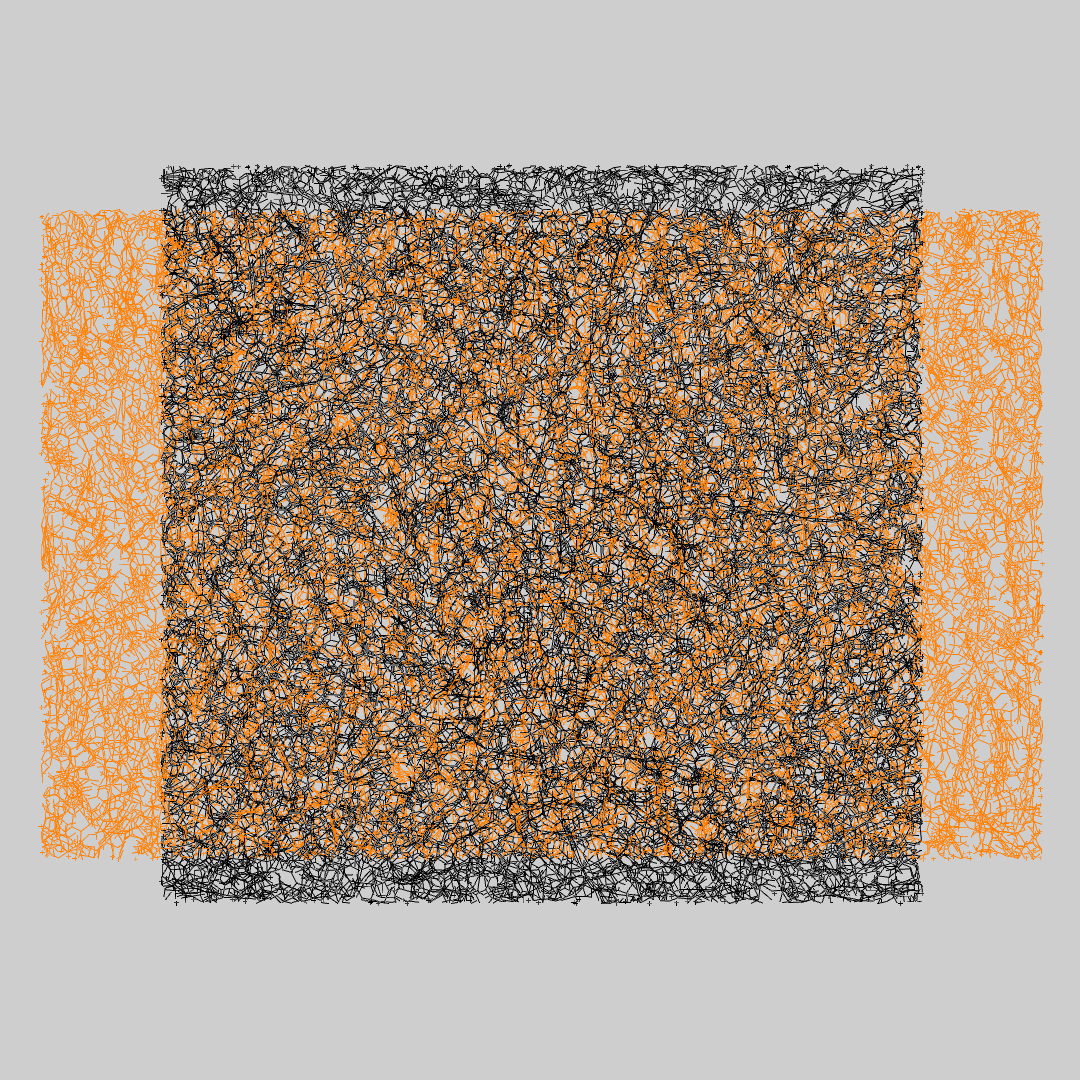}
\includegraphics[width=0.4\columnwidth]{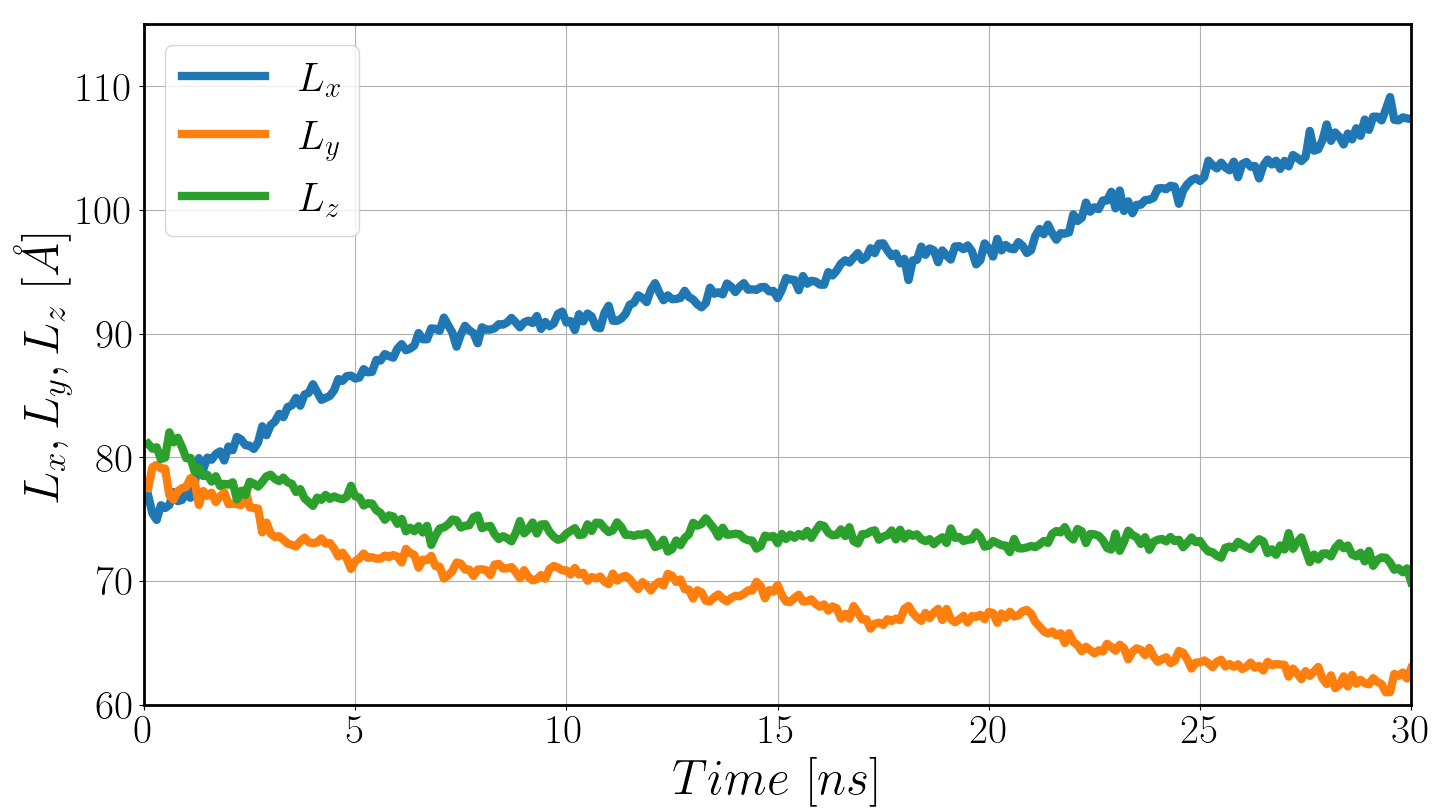}
\includegraphics[width=0.25\columnwidth]{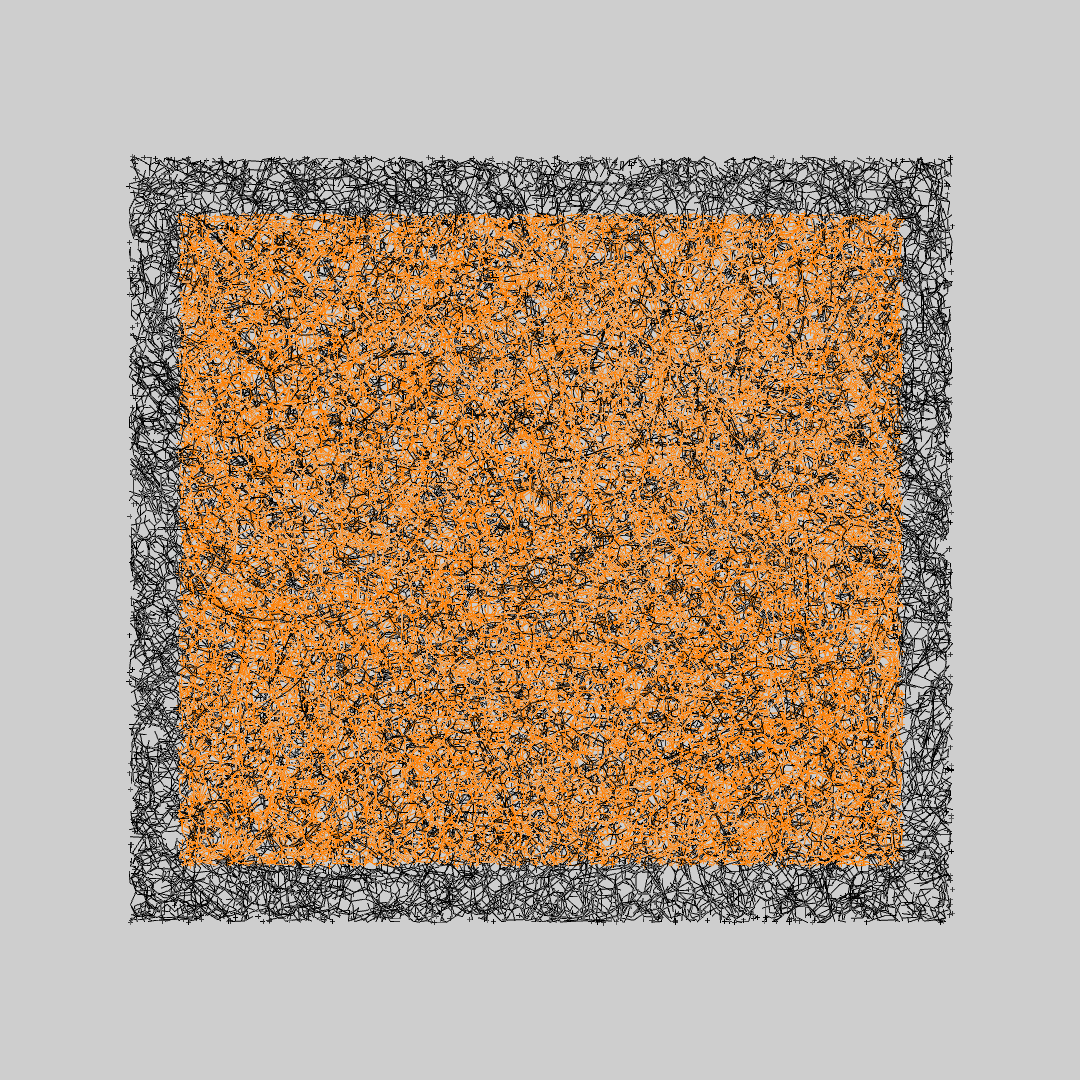}

	\caption{(Top) The snapshots of an acrylate oligomer with 20 azo-chromophores: before irradiation, after 1ns and 30ns of irradiation. (Bottom) The dimensions of volume element: $L_x$ along $\tens E$ and $Ly, Lz$ perpendicular to $\tens E$ (center). The superimposed snapshots of simulation box before and after 30 ns irradiation: xz-projection on the left and yz-projection on the right.   $T=550$ K and $V_0=15$ kcal/mol. } 
	\label{MDsim}
\end{figure}

Consider first the temperatures above the glass transition, for example $T=550$ K (central plot in Fig.\,\ref{MDsim}). The dimension of volume element $L_x$ along the light polarization $\vec E$ first decreases from $80$ nm (the equilibrium value) to $75$ nm, after which it continuously grows, reaching $110$ nm after 30 ns of the irradiation. The dimensions $L_y$ and $L_z$ perpendicular to $\vec E$ decrease. The snapshots of the simulation box before and under irradiation confirm a uniaxial type of light-induced deformation. A slight contraction of the box in the polarization direction occurs between times $t_1$ and $t_2\sim 1$ns, when the order parameter of azobenzenes $S_{azo}$ decreases exponentially. Comparing the snapshots of an azo-oligomer at equilibrium and after $1$ns of irradiation (Fig.\,\ref{MDsim} top), a noticable compactization of the molecule along $\vec E$ is noted. This compactization happens due to alignment of azobenzenes in the plane perpendicular to $\vec E$. The polymer backbone appears to keep its shape till the same moment $t_2$. However, after that, when the orientation state of azo-chromophores does not change much, the polymer backbone begins to align and slightly stretch itself along $\vec E$. Such behaviour has been predicted by analytical theory \cite{Toshchevikov2009} and also observed in coarse-grained simulations \cite{Ilnytskyi2019}.  

\begin{figure}[t]
    \begin{center}
    (a) \includegraphics[width=0.6\columnwidth]{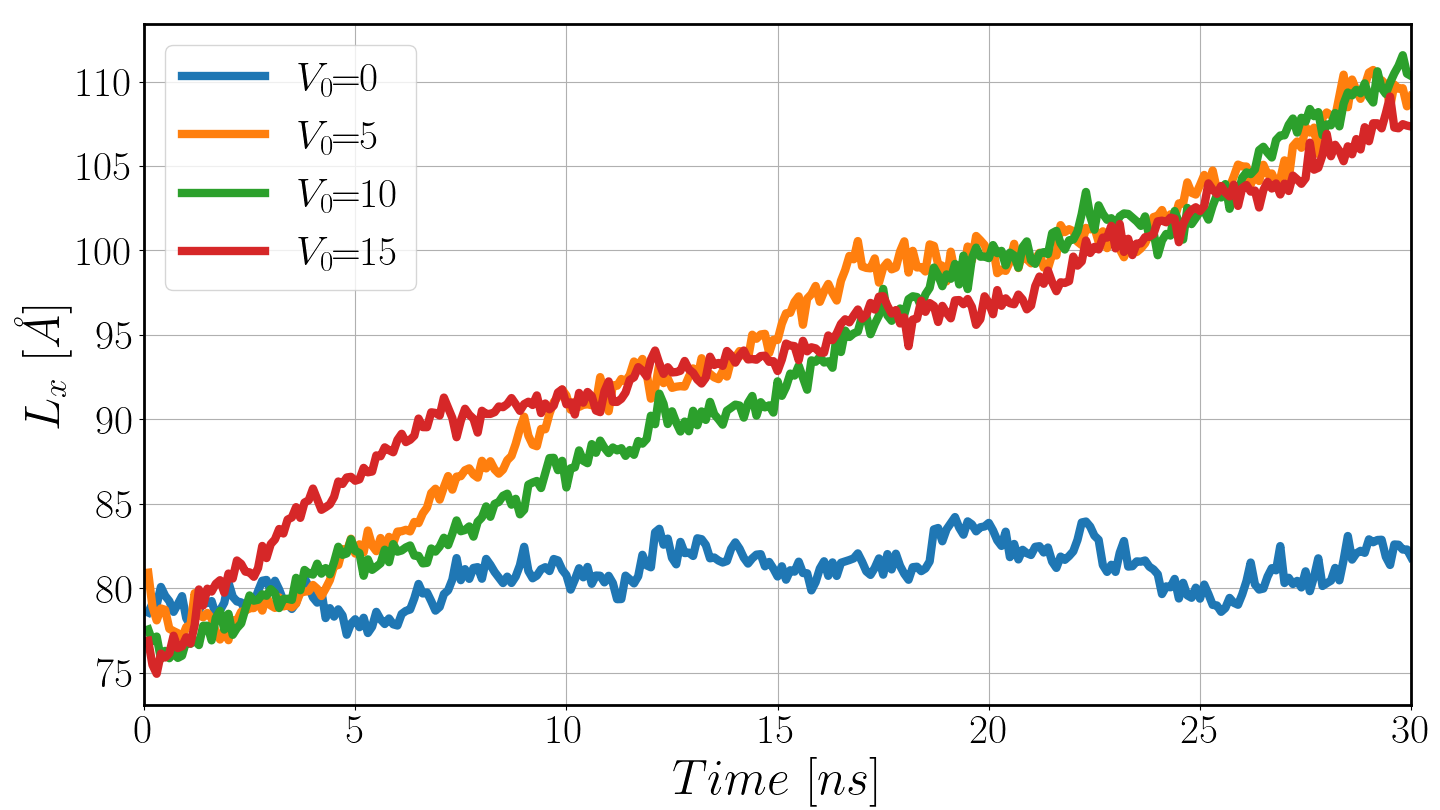}\\
    (b) \includegraphics[width=0.6\columnwidth]{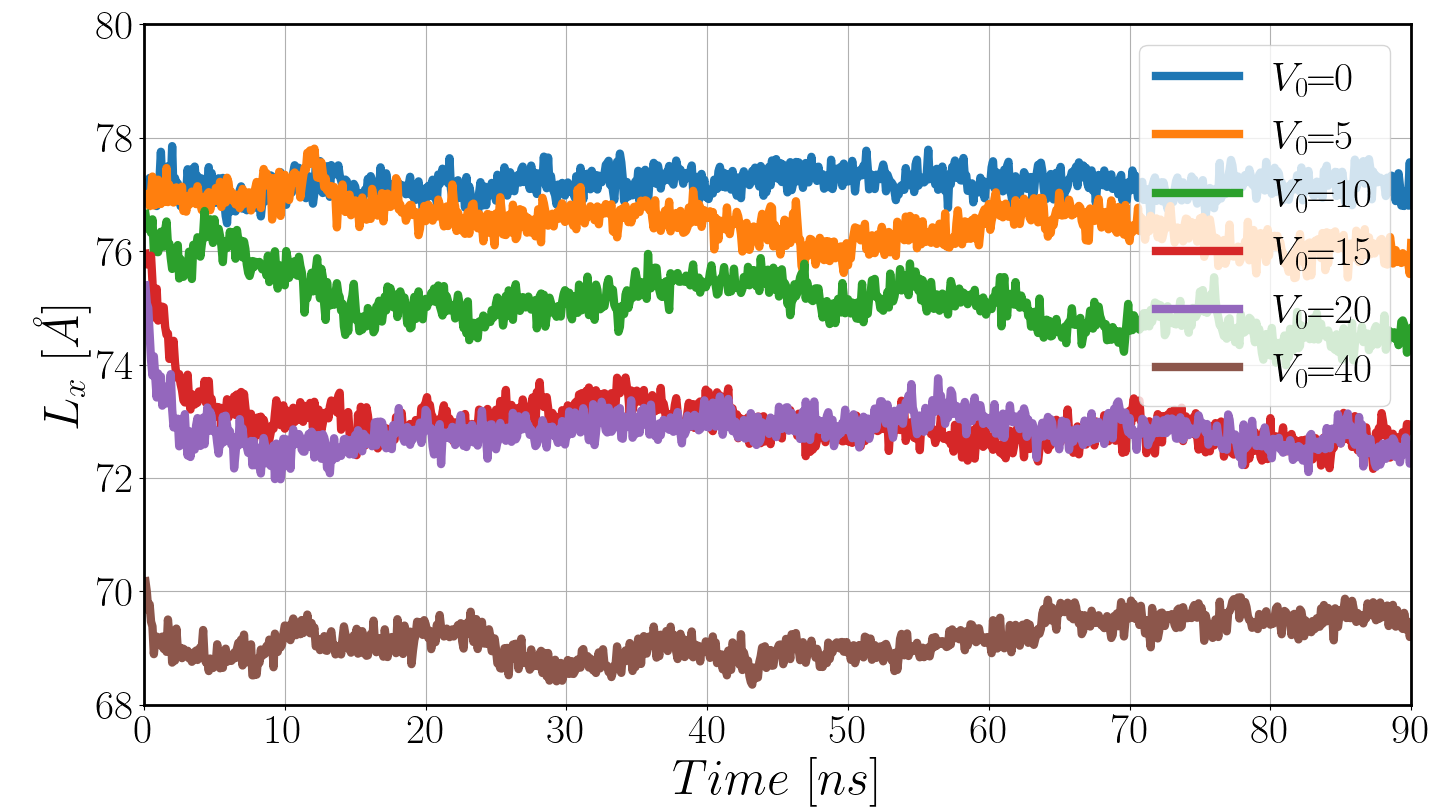}
    \end{center}
    \caption{The dimension of volume element $L_x$ along the polarization direction $\tens E$ for $N_{mol}=20$ and different light-induced torques $V_0$. (a) Slight contraction followed by stretching at $T=550$ K, (b) contraction at $T=400$ K.}
    \label{L_azo20_Tfix_V0}
    \vskip 1cm
\end{figure}

Interestingly, the light-induced torques $5 \le V_0 \le 15$ kcal/mol cause comparable elongation of the volume element after $30$ ns of irradiation at $T=550$ K, only the duration of initial contraction regime decreases with the increase of $V_0$ (Fig.\,\ref{L_azo20_Tfix_V0}a). However, at low temperatures $T<T_g$, only contraction of the volume element along $\vec E$ is observed (Figs.\,\ref{L_azo20_T400_F20} and \ref{L_azo20_Tfix_V0}b). $L_x$ decreases considerably with the increase of light-induced torque $V_0$. This happens at the beginning of irradiation, when the order parameter magnitude grows exponentially with time (Fig.\,\ref{S_azo20_T400}). At larger times, there is no change in the dimensions of the simulation box. The stretching of volume element is not observed, because the conformations of polymer backbones stay frozen at $T<T_g$, as we discuss in the next section.


\subsection{Alignment of polymer backbones}
\label{sec:coupling}

\begin{figure}[h]
    (a) \centering\includegraphics[width=0.6\columnwidth]{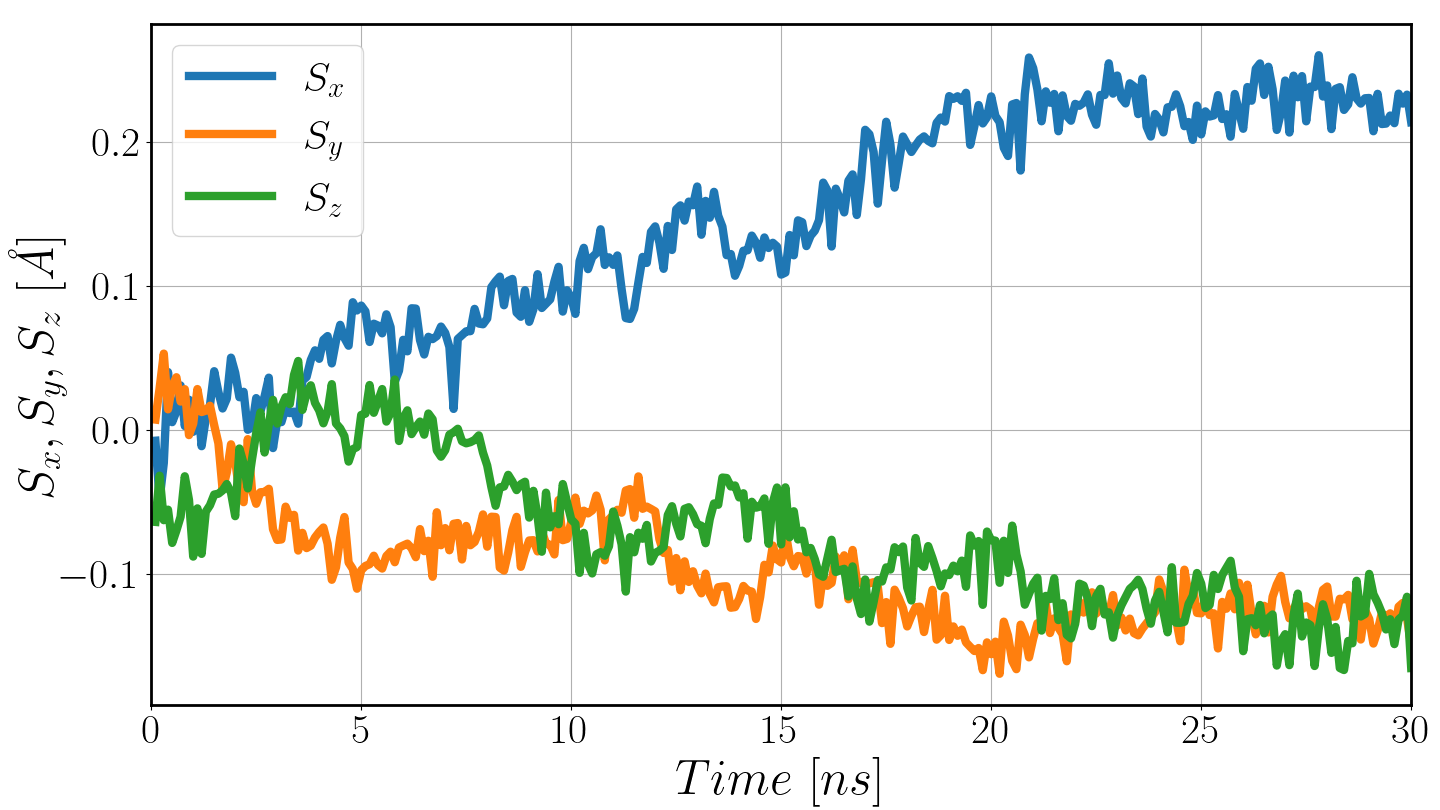}\\
    (b)
\centering\includegraphics[width=0.6\columnwidth]{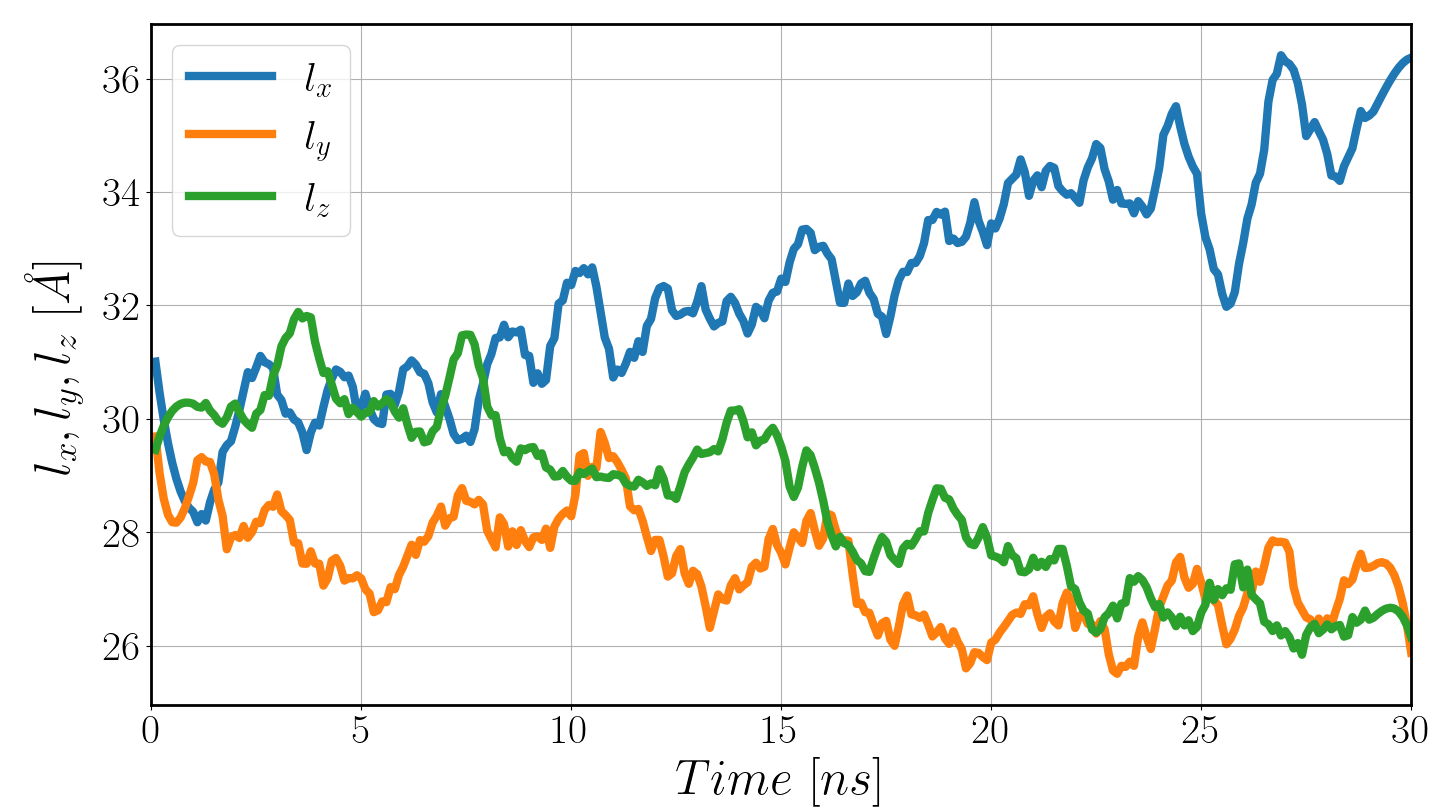}
    \caption{(a) Orientational order parameters $S_x,S_y,S_z$ of oligomer backbones and (b) average backbone sizes $l_x,l_y,l_z$ for $N_{mol}=20$ at $T=550$ K and $V_0=10$ kcal/mol.}
    \label{Backbone_azo20_T550_F10}
\end{figure}

Let us analyze the origin of the long-time stretching of the volume element at elevated temperatures $T>T_g$. The orientation of oligomer backbones with respect to light polarization is described by the order parameter $S_x$, see Eq.~\eqref{S_i}. Its magnitude increases from the equilibrium value 0 to about 0.25 for $T=550$ K and a light-induced torque $V_0=10$ kcal/mol (Fig.\,\ref{Backbone_azo20_T550_F10}a). As expected, the order parameters $S_y$ and $S_z$ take negative values, since the sum of $S_i$ should be zero. Although both $S_y$ and $S_z$ fluctuate noticeably, they reach a similar steady-state value of $-0.12\pm 0.02$, indicating a uniaxial orientational order. The average sizes of the oligomer backbones also show large fluctuations, but the trend is clear. The backbones align along the polarization direction x, resulting in an increase in $l_x$ and a decrease in $l_y$ and $l_z$ (Fig.\,\ref{Backbone_azo20_T550_F10}b). Above the glass transition, the conformations of the polymer backbones are not frozen and can change under irradiation. However, estimates show that the average length of the oligomer backbones fluctuates around 52.5 nm without any stretching or contraction. From this we can conclude that the origin of directional photodeformations above the glass transition is the alignment of backbones along the light polarization. 

In the glassy state, at $T<T_g$, we found no evidence of alignment or deformation of the oligomer backbones even after $90$ ns of simulation time. Both the order parameters and the average sizes fluctuate around their equilibrium values (Fig.\,\ref{Backbone_azo20_T400_F10}). To elucidate the differences in behavior below and above the glass transition, we examined how a light-induced torque influences the orientation distribution of azo-chromophores around the oligomer backbones. 
This distribution is defined by the polar angle $\alpha$, measured as described in Subsection 3.1. The corresponding histograms at various temperatures and time points are presented in Fig.\,\ref{alpha_q}. They are characterized by the average polar angle $\langle\alpha\rangle$ and the shape factor
    \begin{equation}
    \label{shape_factor}
    q=\left<\frac{3}{2}\cos^2 \alpha-\frac{1}{2}\right>.
    \end{equation}
Above the glass transition, at $T=550$ K, $\langle\alpha\rangle$ is approximately $53^{\circ}$ in the dark and decreases only slightly upon applying a light-induced torque of $V_0=10$ kcal/mol for 30 ns, while the shape factor increases from 0.10 to 0.13 over the same period. A similar trend is observed below the glass transition at $T=400$ K for the same torque value. At a stronger torque of $V_0=40$ kcal/mol, $\langle\alpha\rangle$ decreases to $48^{\circ}$ within the first nanosecond of irradiation, and the shape factor increases from 0.08 to 0.20. Extending the irradiation to 90 ns leads to only marginal additional changes in both quantities. 

Positive values of the shape factor indicate contraction of an azo-polymer when the side chains are rigidly coupled to the main chain \cite{Toshchevikov2009,saphiannikova2013nanoscopic}. In contrast, above the glass transition we observe elongation of the azo-polyacrylate. This discrepancy suggests that the flexible spacer plays a crucial role in mediating the transfer of light-induced torque from the azo-chromophores to the oligomer backbones. We investigate this effect in more detail in the following subsection.

\subsection{The role of spacer in torque transfer}
\label{sec:coupling}

\begin{figure}[h]
    \centering\includegraphics[width=0.6\columnwidth]{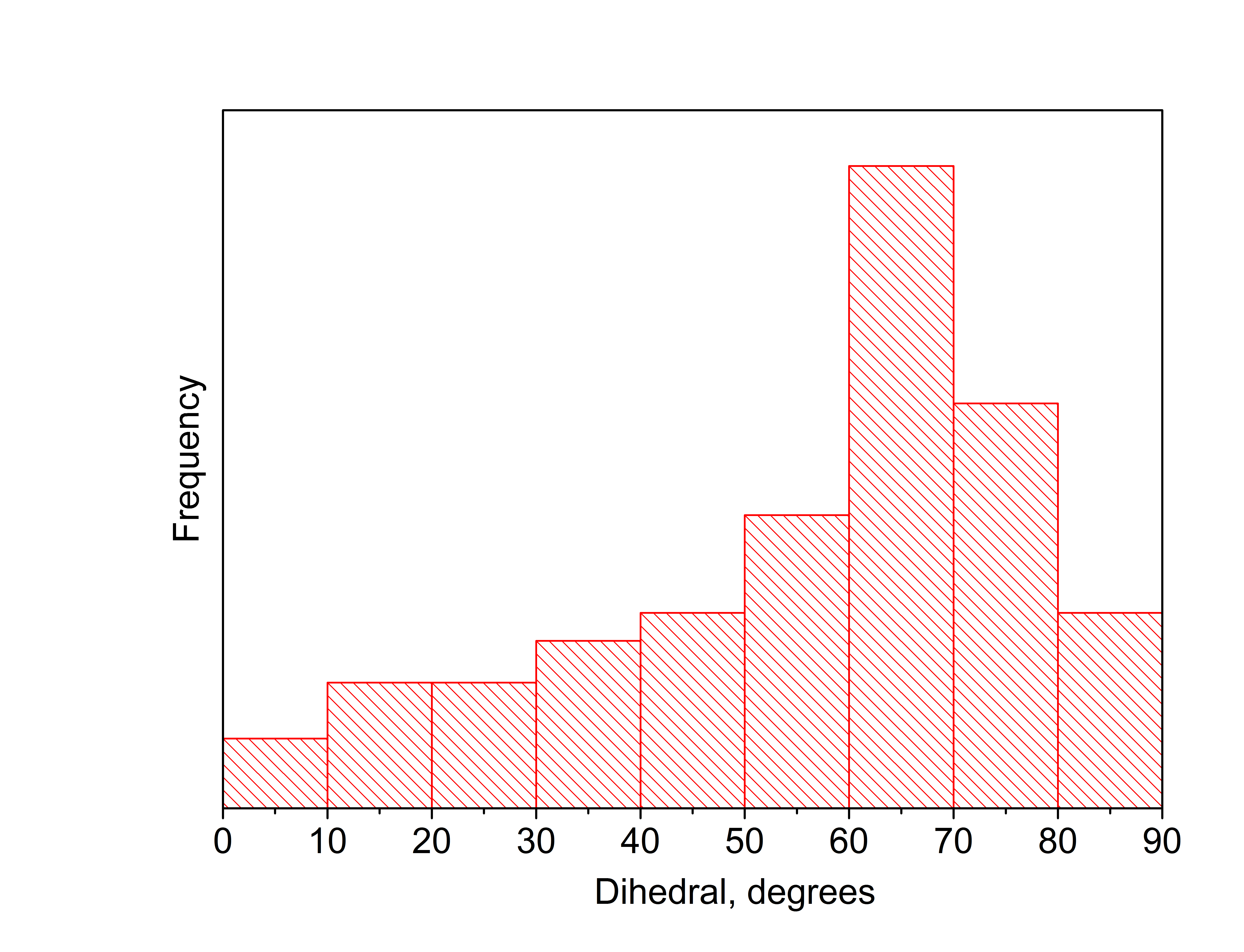}
    \caption{Distribution of dihedral angles between the azobenzene orientation vector \textbf{\textit{u}} and the local backbone direction (C–C vector adjacent to the grafting site). The broad maximum near 55-70$^\circ$ indicates preferentially tilted side-chain orientations, reflecting partial but not rigid coupling between the chromophore and the polymer backbone. This analysis is performed for the system with \textit{N$_{mol}$}=20, \textit{n}=48 chains, simulated at \textit{T}=550 K for 30 ns in dark.}
    \label{distribution1}
\end{figure}

The distribution of dihedral angles between the azobenzene orientation vector \textbf{\textit{u}} and the local backbone direction (defined by the C–C vector preceding and following the grafting point) is shown in Fig.\,\ref{distribution1}. The histogram reveals that the side chains do not populate all rotational states equally but instead exhibit a broad maximum between approximately 55$^\circ$ and 70$^\circ$, with a secondary population extending toward 70-80$^\circ$. This indicates that, although thermally accessible, fully aligned (0-20$^\circ$) or fully orthogonal (80-90$^\circ$) conformations are comparatively rare. The preferred dihedral range around ca. 60$^\circ$ reflects the geometric balance between steric constraints imposed by the grafting site and the inherent stiffness of the azo-chromophore, which tends to orient at an intermediate angle relative to the local backbone segment. In the context of our simulations, this distribution quantifies the baseline angular freedom of azo side chains prior to the application of the light-induced orientation potential. The fact that the ensemble is already biased toward moderately tilted conformations is consistent with the structural analysis presented above: the coupling between side chains and the backbone is neither rigid nor negligible, but instead allows partial torque transfer - a prerequisite for the backbone alignment observed at temperatures above $T_g$. Thus, the dihedral-angle distribution provides a microscopic descriptor of the side-chain flexibility that governs how efficiently chromophore reorientation can drive backbone deformation under illumination.

The chemical linkage between the azo-mesogen and the polymer backbone (illustrated in Fig.~\ref{azo-monomer}b) incorporates an alkyloxy/ester spacer. This arrangement imparts a relatively high degree of conformational flexibility, as the spacer allows rotational and bending motion of the chromophore relative to the backbone segment. Above glass transition temperature, we observe that the spacer frequently undergoes rotations around C–O and also around the aliphatic C–C bonds, giving rise to a rich ensemble of geometries ranging from extended, nearly linear all-trans configurations to gauge- and cis-states. These conformations arise from internal rotations within the flexible alkyloxy–ester segment. Similar behavior has been reported in both experimental and computational studies of azo side-chain liquid-crystalline polymers and azo-polyacrylates, where the internal rotations of the spacer are known to generate broad conformational distributions and significantly influence mesogen mobility, side-chain–backbone coupling, and photomechanical response \cite{Azoreview2024}. The presence of multiple torsional degrees of freedom in the C–C and C–O bonds allows the spacer to sample energetically accessible folded states on nanosecond timescales, effectively modulating the instantaneous orientation of the chromophore relative to the backbone. This multibond rotational flexibility is therefore a key structural feature that enables efficient torque transfer from the light-driven azo unit to the polymer main chain while still permitting substantial local relaxation of the side chain. It also helps explain the broad dihedral-angle distributions observed in our analysis (Fig.\,\ref{distribution1}), as the backbone–chromophore connection is not restricted to a single rotational axis but behaves as a highly compliant, thermally active joint. 

To quantify the conformational flexibility of the spacer connecting the azo side group to the polymer backbone, we compared the energy of a fully extended, optimized all-trans reference structure with the computed energies of conformations extracted from the molecular dynamics trajectory for the system with \textit{N$_{mol}$}=20, \textit{n}=48 chains, simulated at \textit{T}=550 K for 30 ns in dark. In all the calculations, density functional theory is used as implemented in Gaussian \cite{g09} with Perdew-Burke- Ernzerhof functional and 6-31G* basis set. In azo-polymer, the chromophore is linked to the main chain via a spacer (Fig.~\ref{azo-monomer}b), which provides several torsional degrees of freedom with hindered but accessible rotation. The conformations sampled in the MD run are therefore not strictly all-trans: they include various gauche and, in some cases, cis-like arrangements along the spacer, as well as a slight loss of planarity of the trans-azobenzene core. As a consequence, the single-point energies of the MD snapshots are systematically higher than that of the optimized all-trans minimum, with typical excess energies on the order of ca. 60 kcal/mol. This value should not be interpreted as the barrier of a single dihedral, but rather as the cumulative energetic penalty associated with simultaneously exciting several torsions and distorting the azobenzene $\pi$-system away from its near-planar ground-state geometry. Similar sensitivity of azo-derivatives to phenyl-ring torsion and local environment has been highlighted in quantum-chemical studies of isolated azobenzenes and their substituted analogues, where the trans form is found to be (nearly) planar and separated from twisted conformers by only shallow potentials, while the overall cis–trans gap remains large.

\section{Conclusions}
\label{sec:conclusions}

Fully atomistic MD simulations of dense azo-polyacrylate samples allow us now to answer on the questions posed in the introduction to this study:
\begin{enumerate}
\item It is possible to observe the orientation of acrylate backbones along the polarization direction above the glass transition temperature. The simulation box first slightly contracts and then elongates in this direction.   
\item The contraction of the box correlates with the rapid orientation of azobenzenes, followed by much slower orientation of main chains. The separation between these two time scales increases dramatically, when the governing parameter temperature drops below the glass transition.
\item The length of the main chains does not affect the glass transition temperature and the light-induced processes for simulated oligomers. This is due to the absence of entanglements in these short but densely grafted azo-polymer structures. 
\end{enumerate}

On one side, we obtained a very positive result, that fully atomistic MD simulations allow us to observe the orientation of acrylate backbones along the polarization direction, as predicted by analytical theories. On the other hand, our studies become hampered by extremely
short observation times. This is an intrinsic limitation of all-atom MD simulations.
Another valuable result is a better understanding of the nature of coupling between the orientation of photosensitive azobenzene groups and the backbones in
a flexible polymer. 

Indeed, in contrast to rigid azo-polyesters, the azo-polyacrylate examined here contain an alkyloxy–ester spacer that possesses several accessible torsional degrees of freedom. The MD trajectories reveal broad dihedral-angle distributions and frequent sampling of gauche and partially folded states relative to an all-trans reference. This conformational richness allows the spacer to absorb a substantial part of the light-induced torque through local relaxation. As a result, the effective coupling between the chromophore and the backbone is significantly weaker than in architectures with short and rigid linkers. The torque imparted by photoaligned azobenzenes is therefore transmitted to the main chain only partially and with delay, becoming efficient only when the thermal energy available above glass transition temperature permits the backbone to respond. This mechanistic picture explains why highly flexible azo-polyacrylates still display robust photodeformations experimentally: while the side-chain/backbone coupling is attenuated, it is not eliminated, and its action is unlocked when chain mobility is sufficient.

\noindent\textbf{Acknowledgments}\\
Financial support from Deutsche Forschungsgemeinschaft (DFG) under
grant GR 3725/10-1 is greatly appreciated. We would like to acknowledge the high-performance computing support from
the Center for Information Services and High Performance Computing (ZIH) provided by Technische
Universit\"at Dresden. The authors acknowledge the valuable contribution of Dr. Markus Koch, who developed and integrated an effective orientation potential~\cite{Toshchevikov2014,Toshchevikov2017} into the simulation code LAMMPS (https://github.com/Markus91Koch/LAMMPS{\_}Extensions) as part of his Ph.D. research, forming the methodological basis for the present study.

\bibliographystyle{is-unsrt}
\bibliography{atomazo.bib}

\newpage
\section*{\LARGE Supporting Information}
\label{sec:SI}

\renewcommand{\thefigure}{S\arabic{figure}}
\setcounter{figure}{0}

\begin{figure}[h]

\centering\includegraphics[width=0.6\columnwidth]{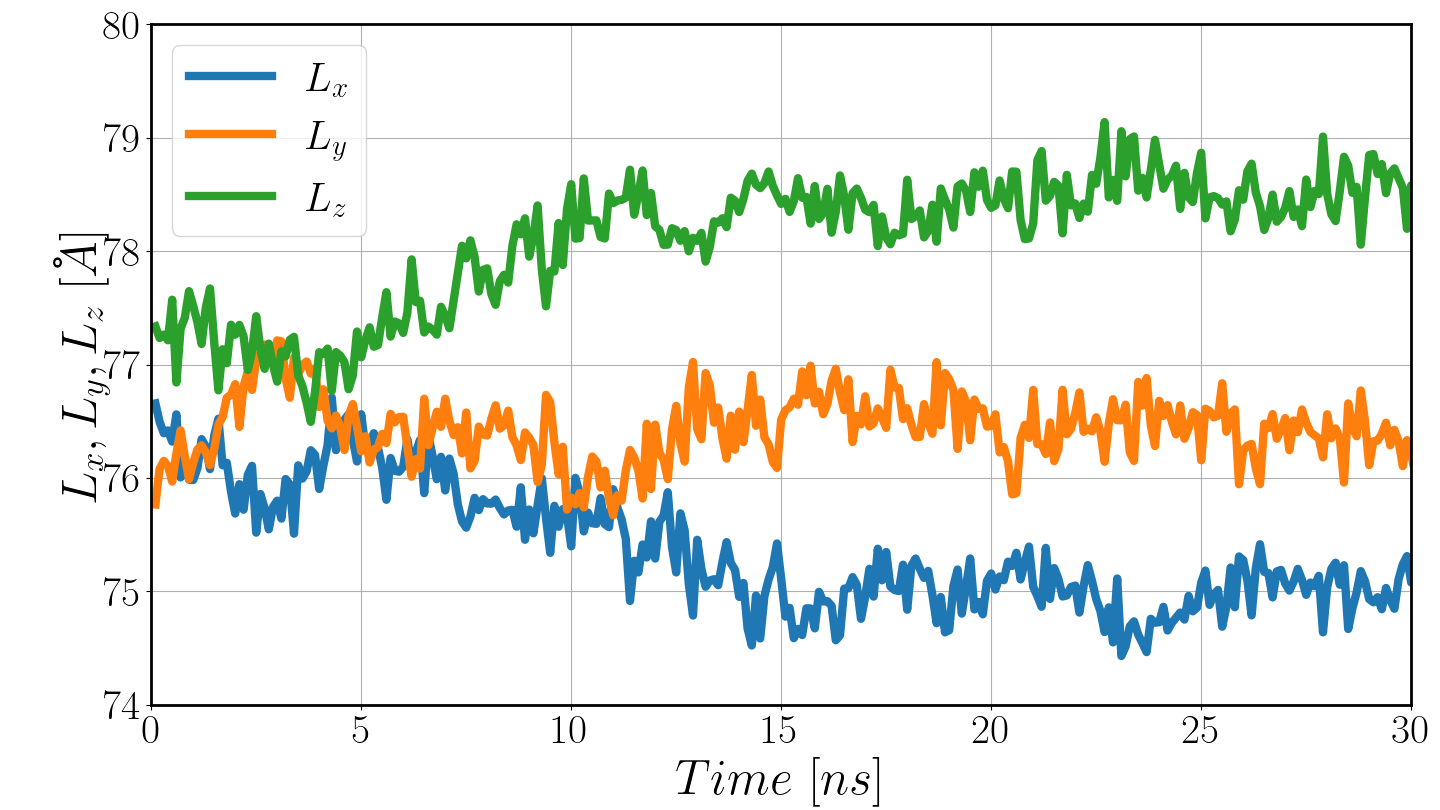}
    \caption{The dimensions of volume element: Lx along E and Ly, Lz perpendicular to E for $N_{mol}=20$ at $T=400$ K and $V_0=10$ kcal/mol.}
    \label{L_azo20_T400_F20}
\end{figure}

\begin{figure}[h!]
    (a) \centering\includegraphics[width=0.6\columnwidth]{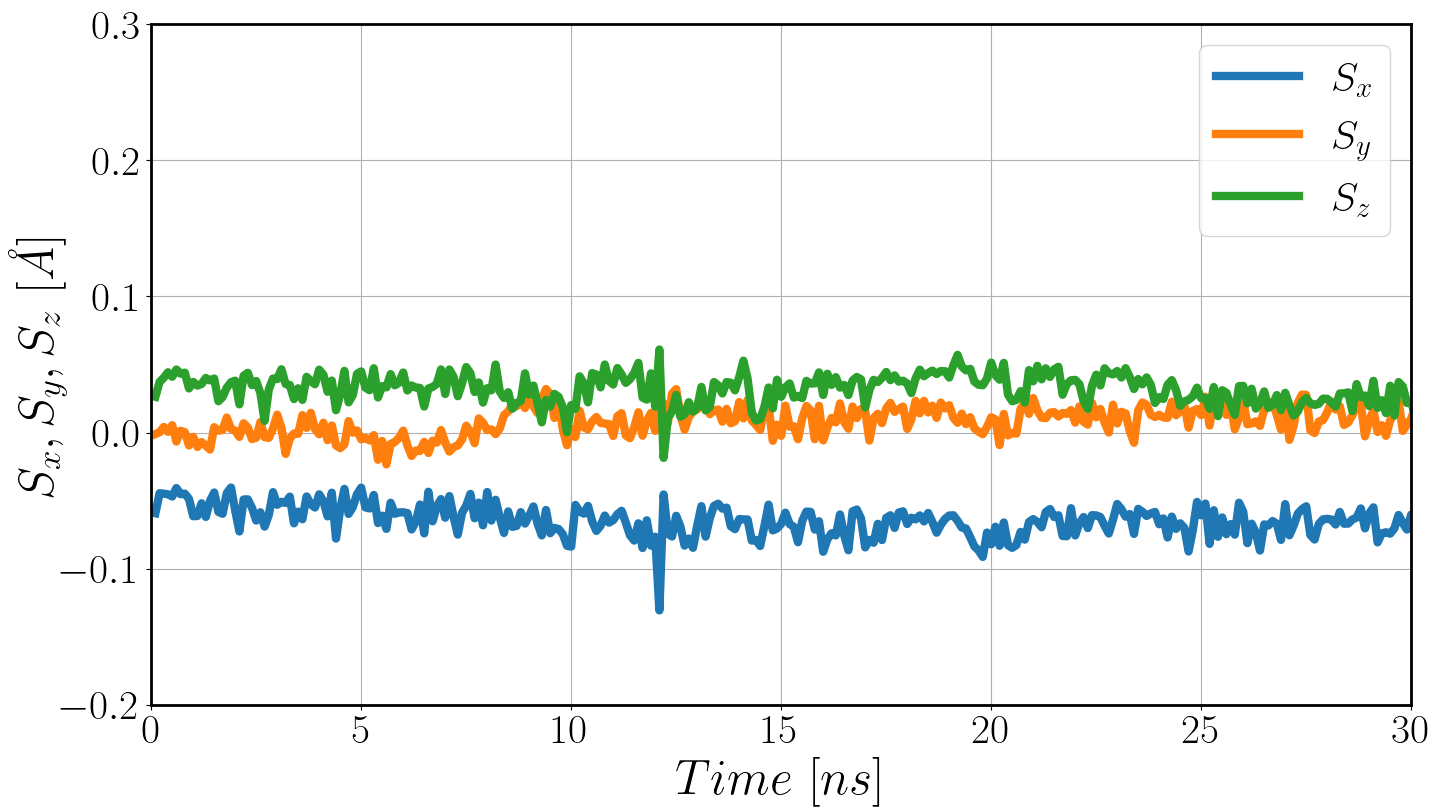}\\
    (b)
\centering\includegraphics[width=0.6\columnwidth]{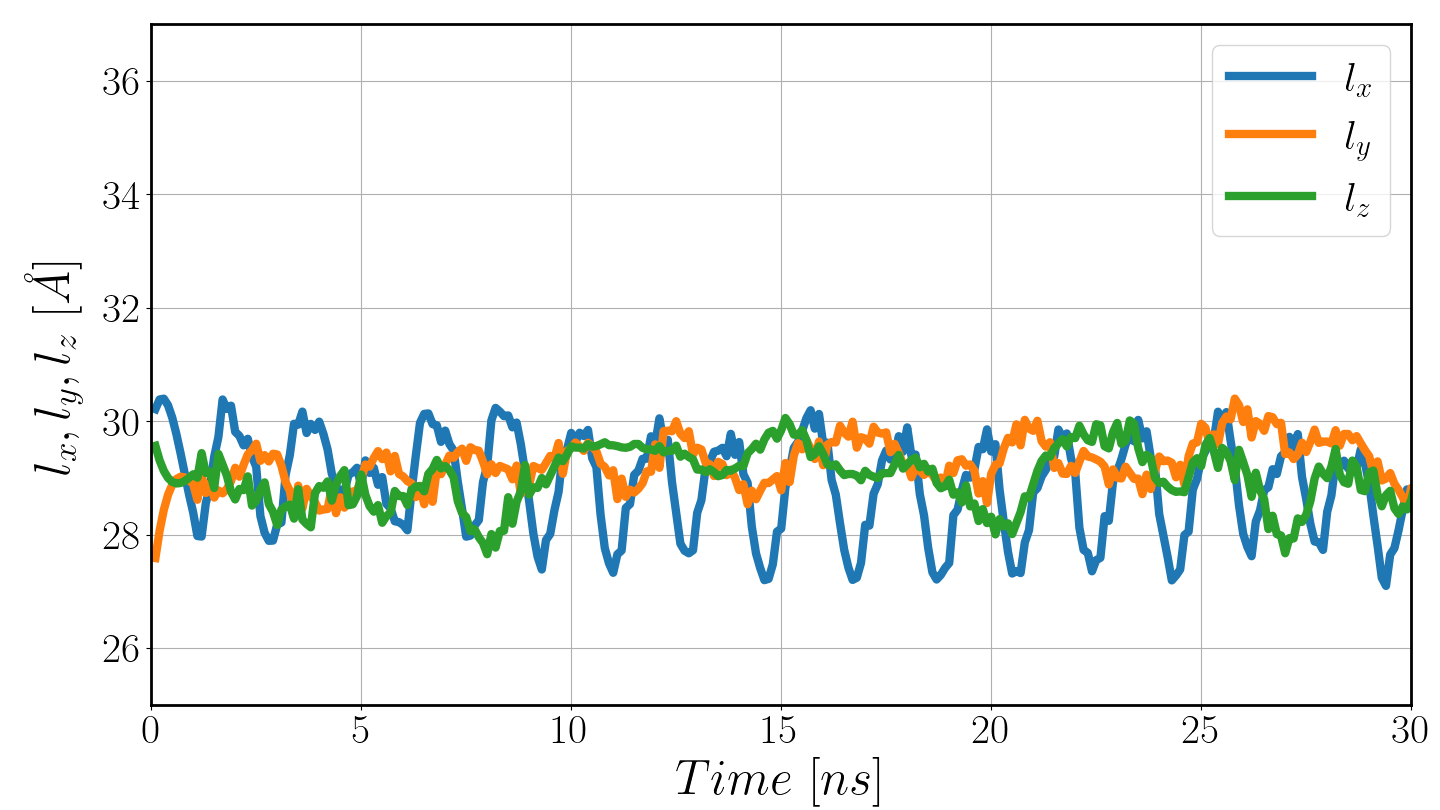}
40    \caption{(a) Orientational order parameters $S_x,S_y,S_z$ of oligomer backbones and (b) average backbone sizes $l_x,l_y,l_z$ for $N_{mol}=20$ at $T=400$ K and $V_0=10$ kcal/mol.}
    \label{Backbone_azo20_T400_F10}
\end{figure}

\includepdf[scale=0.85, nup=1x3,page=-]{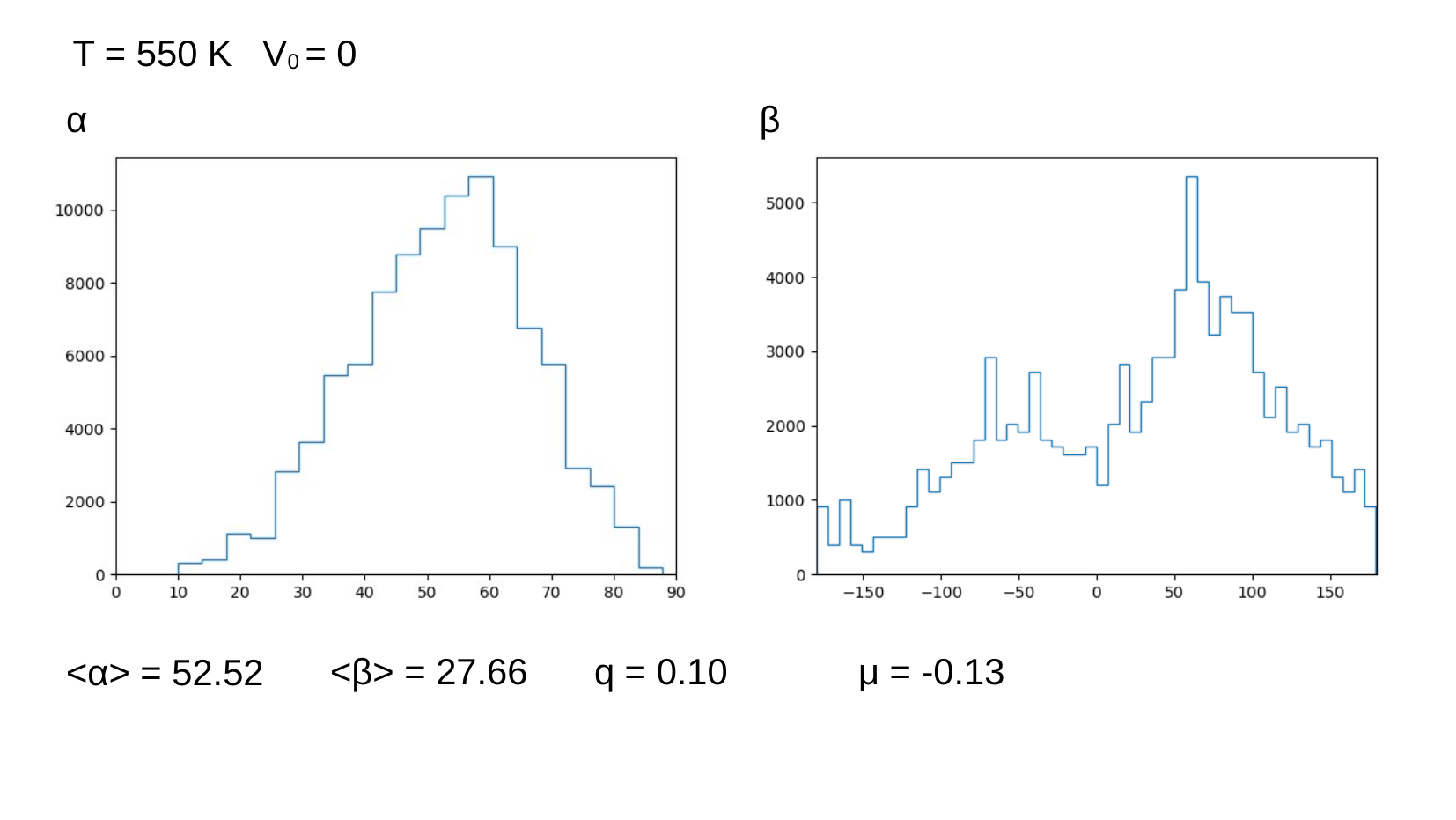}

\begin{figure}[h!]
\label{alpha_q}
\end{figure}

\end{document}